\documentclass[12pt,preprint]{aastex}
\slugcomment{Submitted to the {\it Astrophysical Journal}}
\shorttitle{Extragalactic FIR Background}
\shortauthors{Chary \& Pope}
\newcommand{\cirbu}{nW\,m$^{-2}$\,sr$^{-1}$}
\newcommand{\sfru}{M$_{\sun}$ yr$^{-1}$ Mpc$^{-3}$}

\begin{document}

\title{New Observational Constraints and Modeling of the Infrared Background: Dust Obscured
Star-Formation at $z>1$ and Dust in the Outer Solar System}

\author{Ranga-Ram Chary\altaffilmark{1} \& Alexandra Pope\altaffilmark{2,3}}
\altaffiltext{1}{{\it Spitzer} Science Center, California Institute of Technology, Pasadena, CA 91125; {\tt rchary@caltech.edu}}
\altaffiltext{2}{National Optical Astronomy Observatories, Tucson, AZ 85719}
\altaffiltext{3}{Spitzer Fellow}

\begin{abstract}

We provide measurements of the integrated galaxy light at 70, 160, 250, 350 and 500\,$\mu$m using deep
far-infrared and submillimeter data from space ({\it Spitzer}) and balloon platform (BLAST)
extragalactic surveys. We use the technique
of stacking at the positions of 24\,$\mu$m sources, to supplement the fraction
of the integrated galaxy light that is directly resolved through direct detections. We demonstrate that
the integrated galaxy light even through stacking,
falls short by factors of 2$-$3 in resolving the extragalactic far-infrared background.
We also show that previous estimates of the integrated galaxy light (IGL) through stacking,
have been biased towards high values. This is primarily due to multiple counting of the far-infrared/submillimeter
flux from 24\,$\mu$m sources 
which are clustered within the large point spread function of a brighter far-infrared source.
Using models for the evolution of the luminosity function at $z<1.2$
which are constrained by observations at 24\,$\mu$m and 70\,$\mu$m, and 
which are consistent with the results from the stacking analysis, 
we find that galaxies
at $z<1.2$, account for $\sim95-55$\% of the extragalactic far-infrared
background in the $\sim70-500\,\mu$m range respectively. This places
strong upper limits on the fraction of dust obscured star-formation at $z>1$, which are remarkably, below
the values derived from the extinction corrected ultraviolet luminosities of galaxies. 
The largest fraction of the total $40-500$\,$\mu$m EBL comes from galaxies between 
L$_{\rm IR}$=L$(8-1000\,\mu m)$=10$^{10-11.5}$\,L$_{\sun}$; ultraluminous infrared galaxies 
with L$_{\rm IR}>10^{12}$\,L$_{\sun}$ contribute only 5\%.
We use the results to
make predictions for the nature of galaxies that extragalactic surveys with {\it Herschel Space Observatory} will reveal.
Finally, from our constraints on the far-infrared IGL, we provide evidence for the existence of ice mantle dust, orbiting
the sun at a distance of $\sim$40 AU, which is contributing intensity to both the near- and far-infrared
background. The presence of this component which is a tiny fraction of the zodiacal light emission even
at high ecliptic latitudes, eliminates any discrepancy between the integrated
galaxy light and the diffuse, extragalactic background light at all infrared wavelengths.

\end{abstract}


\section{Introduction}

The extragalactic background light (EBL)\footnote{Also referred to as the Cosmic Infrared Background} at
far-infrared (FIR) wavelengths, spanning a wavelength range of $40-1000$\,$\mu$m, is the sum total of radiation
from all astrophysical processes over cosmic time that is absorbed by dust and re-radiated. It places
a strong constraint on the fraction of nucleosynthesis and accretion activity that is obscured by dust. Decomposition
of the EBL into the contribution from individual galaxies as a function of redshift,
provides a more complete understanding of the rate of build up of stellar mass and the growth
of black holes over cosmic time and thereby helps identify the epoch at which most of the stars
and black holes in the present day Universe formed.

The FIR EBL is thought to have
a total intensity between $40-1000$\,$\mu$m of $\sim$25\,\cirbu\ although values as high as $\sim$55\,\cirbu\ have been suggested by 
DIRBE observations \citep{Hauser, Finkbeiner:00}. Much 
of the difference arises due to the difficulty in subtracting the contribution from
zodiacal light emission. The zodiacal light
contributes $\sim$30\% of the line of sight intensity at 140$\mu$m in even a relatively
high ecliptic latitude ($\beta\sim45\arcdeg$)
field like the Lockman Hole, with a rapidly increasing fractional contribution at shorter wavelengths \citep{Hauser}.
Even small errors in the subtraction of the zodiacal light propagate through as large errors in the estimate of the
FIR EBL, an issue which has plagued measurements of the EBL at 60\,$\mu$m and 100\,$\mu$m.

Despite the fact that it is measured with a factor of $\sim$100 less precision than the cosmic microwave background, 
the FIR EBL is arguably the second most energetic extragalactic background after the cosmic microwave background which
has a total intensity of $\sim$1000\,\cirbu. The
intensity of the FIR EBL
is comparable, if not in excess of the intensity of the optical/near-infrared extragalactic background, whose measurement
is affected by similar uncertainties associated with the zodiacal light contribution \citep{Bernstein, Levenson:07}.
Since the optical/near-infrared EBL
measures the redshifted contribution of unobscured starlight and AGN activity, integrated over cosmic time, the fact
that the two intensities are comparable, illustrates
the importance of dust obscuration in quantifying the bolometric
luminosity of star-forming galaxies and active galactic nuclei.

Ever since the direct measurement of the EBL using the Far-Infrared Absolute Spectrometer (FIRAS) and the Diffuse
Infrared Background Experiment \citep[DIRBE;][]{Puget, Hauser, Fixsen}, various attempts 
have been made to identify the individual sources contributing to the background. 
The sum of the contribution of individual galaxies is termed the integrated galaxy light (IGL).
Using backward evolution models which fit multiwavelength number counts,
\citet[][hereafter CE01]{CE01} demonstrated that dust obscured infrared luminous galaxies contribute 85\% of the 140\,$\mu$m
EBL and that 82\% of the peak of
the FIR EBL arises from galaxies at $z<1.5$. The 140\,$\mu$m EBL is $\sim$60\% of the total FIR EBL in the CE01 models. 
This implies that about 50\% of the total FIR EBL arises at $z<1.5$.

\citet{Elbaz} utilized model spectral energy distributions to estimate
the contribution to the FIR EBL of infrared luminous galaxies individually
detected by the ISOCAM in a deep 15\,$\mu$m survey in the Hubble Deep Field-North.
They found that the ISOCAM LIRGs at $z<1$ contribute 16$\pm$5\,\cirbu\ of the 140$\mu$m background. 
More recently, by stacking the far-infrared data, \citet{Bethermin} showed that galaxies with 24$\mu$m flux densities $>$35\,$\mu$Jy
contribute 9.0$\pm$1.1\,\cirbu\ of the 160\,$\mu$m background. Based on photometric redshifts, these
galaxies are primarily thought to be infrared luminous galaxies at $z\sim1$ with a tail extending to $\sim2.5$ \citep{Caputi}. 

In this context, the recent results from Balloon Large Aperture Submillimeter Telescope (BLAST) have been 
somewhat surprising \citep{Devlin, Marsden}. Using a technique similar to \citet{Dole}, the BLAST team
estimated that stacking on sources with 24$\,\mu$m flux density greater than 13\,$\mu$Jy
results in an IGL of 8.6$\pm$0.6\,\cirbu\ at 250\,$\mu$m (i.e. the majority of the EBL at this wavelength). 
Furthermore, the conclusion is that 24\,$\mu$m detected galaxies contribute
$\sim$80-90\% of the total FIR EBL between 250 and 500\,$\mu$m.
The deep 24\,$\mu$m data which have been used for their analysis are only sensitive to galaxies brighter than 10$^{10}$\,L$_{\sun}$
at $z\sim1$ and $>$10$^{11}$\,L$_{\sun}$ at $z\sim2$ \citep{Chary:07a}. It is well known that there is a whole population of
Lyman-break galaxies which are less luminous and which appear to have significant dust obscuration based on their
UV-slope. Although they should be contributing to the EBL, the BLAST analysis seems to suggest that the
contribution from fainter galaxies is surprisingly small.

In this paper, we utilize deep, public {\it Spitzer} data taken as part of the Far-Infrared Deep Extragalactic Legacy (FIDEL; P.I. M. Dickinson) survey\footnote{http://data.spitzer.caltech.edu/popular/fidel/} 
at 70 and 160\,$\mu$m
and the publicly released BLAST maps at 250, 350 and 500\,$\mu$m to evaluate the consistency of these different estimates.
We combine direct detections of sources with stacking analysis to provide lower limits to the integrated galaxy light.
We utilize recent measurements of the far-infrared luminosity function out to $z\sim1.3$, 
in conjunction with the 
model spectral energy distribution of galaxies based on various spectroscopic and imaging data,
to determine the nature of
sources contributing to the far-infrared background as a function of redshift and luminosity. We 
place constraints on dust obscured star-formation at $z>1$ and discuss implications for 
deep {\it Herschel} extragalactic surveys. Finally, we propose the existence of a previously unknown cloud of high albedo dust grains
orbiting the Sun at a distance of $\sim$40 AU which contributes significantly
to the intensity of the near- and far-infrared background.

Throughout this paper we assume a standard cosmology with $H_{0}=71\,\rm{km}\,\rm{s}^{-1}\,\rm{Mpc}^{-1}$, $\Omega_{\rm{M}}=0.27$ and $\Omega_{\Lambda}=0.73$.

\section{The Integrated Galaxy Light from Direct Detections and Stacking}

The integrated galaxy light is the sum of the observed light 
from individually detected galaxies at a particular wavelength. Due 
to the sensitivity limit of a survey, it is difficult to directly detect the dominant population of
galaxies contributing at a particular wavelength. This is particularly
difficult at far-infrared wavelengths, since the sensitivity of far-infrared detectors are relatively
poor compared to those at shorter wavelengths.  Furthermore, due to the large point spread function, source 
confusion dominates the limiting flux of a survey.

It is possible to assess the total intensity
contribution from galaxies below the detection limit of a survey using the technique
of stacking. This requires knowing the positions of the galaxies in the survey area from a deeper,
less confused survey. The 24\,$\mu$m observations of the FIDEL fields with {\it Spitzer} provide
a particularly useful sample of galaxies. The surveys at this wavelength are $\gtrsim$150 times
deeper in flux density than a deep far-infrared survey.
By sampling redshifted hot dust and polycyclic aromatic hydrocarbon features, 24\,$\mu$m surveys
reveal a sample of galaxies which have significant dust in them and are likely to be contributing
to the emission at far-infrared wavelengths due to the strong observed correlation between
mid-infrared and far-infrared luminosities of galaxies (CE01). By stacking the far-infrared
data at the positions of these 24\,$\mu$m sources,
the background limited noise $\sigma$ can be reduced by $\sigma$/$\sqrt{N}$ where $N$ is the number
of sources that are stacked on. In reality however, the
flux from the wings of
individual bright sources and from the uncertainty in the subtraction of these brighter
source results in an additional term, such that the noise in the stacked image is:
\begin{equation}
\frac{\sqrt{\sum_{i}{\sigma^2 + {\sigma^{2}_{conf,i}}}}}{N}
\end{equation}
with $\sigma^{2}_{conf,i}$ being the variance in
the residual map in the vicinity of the i$^{th}$ source after it has been fit and subtracted.
If the sources are not subtracted from the map, then $\sigma_{conf,i}$ is simply the Poisson noise
from the flux of the i$^{th}$ source which is typically larger than the variance in the residual map.
Similarly, the extended wings of the point spread function results in the stacked flux biased high as we will demonstrate
in the next section.
Thus, to avoid contamination of the stacked flux, stacking should generally be undertaken on maps in which 
the individual sources detected
with high significance are fit for and subtracted.

Since stacking can only reveal average properties of the stacked sample of galaxies, selecting a homogeneous
sample of objects provide more illustrative
results than stacking a heterogeneous sample such as those selected
at optical/near-infrared wavelengths. For stacking at far-infrared wavelengths, 24\,$\mu$m-selected sources are the preferred choice due to their superior spatial resolution, depth and stable image quality compared to longer wavelengths \citep{Rieke04}.

Furthermore, extreme care must be taken to avoid counting the flux in the wings
of individual sources multiple times, an effect which is important at far-infrared wavelengths with the
large point spread function. Care must also be taken to include the clustering properties of the stacked
sources in the error analysis which otherwise results in underestimates of the uncertainty in the stacked
flux. 

\subsection{{\it Spitzer} Far-Infrared Surveys}

We use three extragalactic fields for our IGL analysis: GOODS-N, ECDFS and Extended Groth Strip (EGS); all 
of which were observed at 24$\,\mu$m as part of the GOODS
{\it Spitzer} and/or FIDEL
Legacy surveys (P.I. Mark Dickinson). The depths of the Spitzer 24$\,\mu$m data 
varies from field-to-field. For consistency in our stacking analyses, we chose a 24$\,\mu$m depth for 
each field such that the completeness limits were comparable. These flux limits 
are listed in Table \ref{tab:blast}. The 24$\,\mu$m source catalogs are constructed using the positions
of sources detected at shorter wavelengths ($3-8$\,$\mu$m) as priors \citep{Magnelli}.

\subsubsection{70\,$\mu$m}

We utilize a mosaic constructed using
the public 70\,$\mu$m observations of the GOODS-N field taken as part of the FIDEL program 
in conjunction with
deep observations described in \citet{Frayer}. Since the combined mosaic covers the entire
GOODS-N area, there is corresponding deep 24\,$\mu$m coverage over this mosaic.
The total field size with deep coverage is 174.3\,arcmin$^{2}$.
The {\it Spitzer} beam size at 70$\mu$m is 18$\arcsec$ Full Width at Half Maximum
(FWHM) while the beam size at 24$\mu$m is 5.7$\arcsec$ FWHM (Table \ref{tab:blast}). 
At this spatial resolution, almost all extragalactic
sources are point sources.
Using positional priors from the much deeper 24\,$\mu$m data, \citet{Magnelli}
have used point source fitting to estimate the 70\,$\mu$m
flux density of each 24\,$\mu$m source. Using extensive simulations, they have characterized the
reliability and completeness of the FIDEL 70\,$\mu$m survey. The 80\% completeness limit of the
70\,$\mu$m survey is 3 mJy although individual sources in unconfused regions can be detected down
to fainter flux densities. 

\begin{figure}
\plottwo{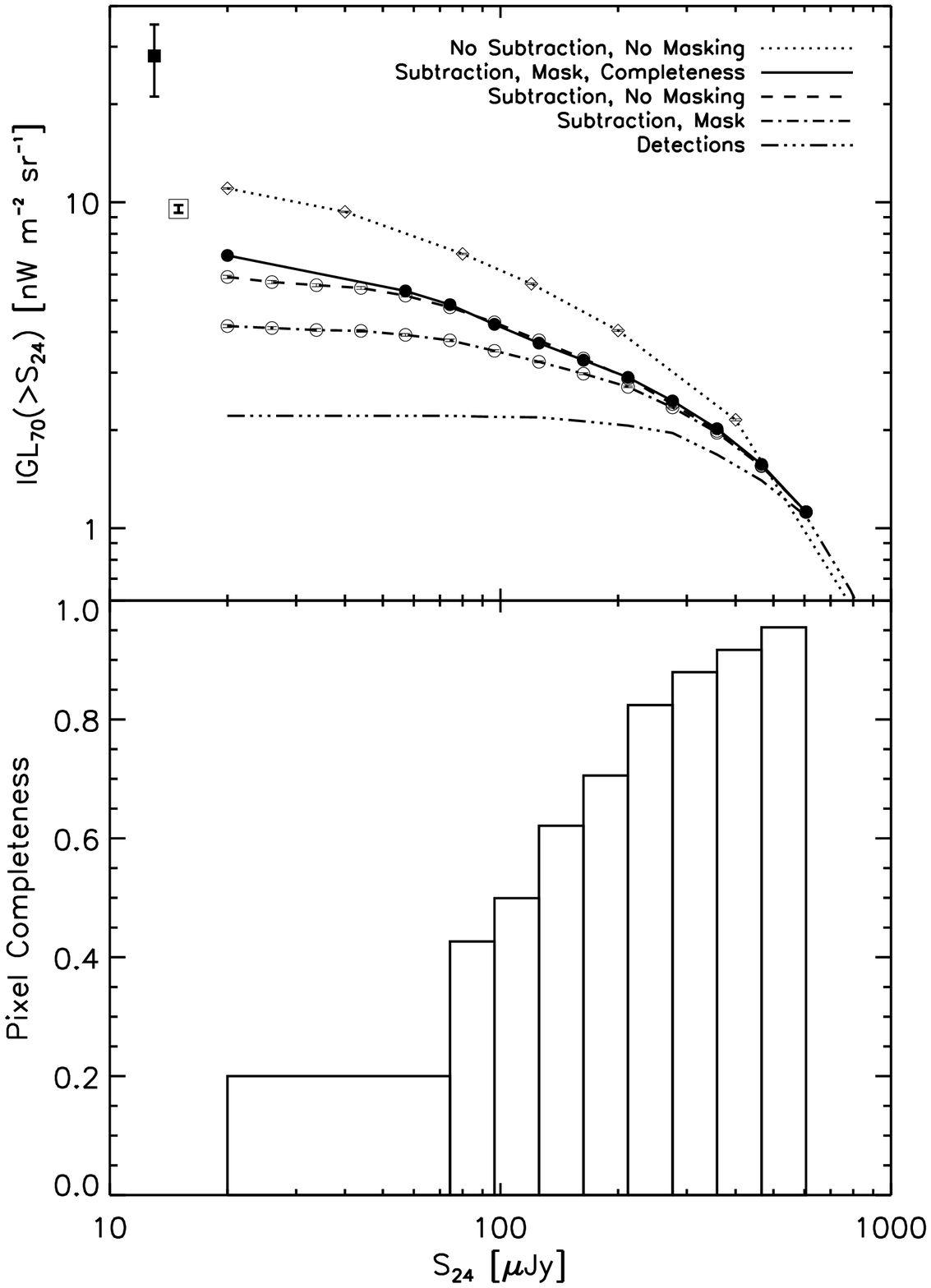}{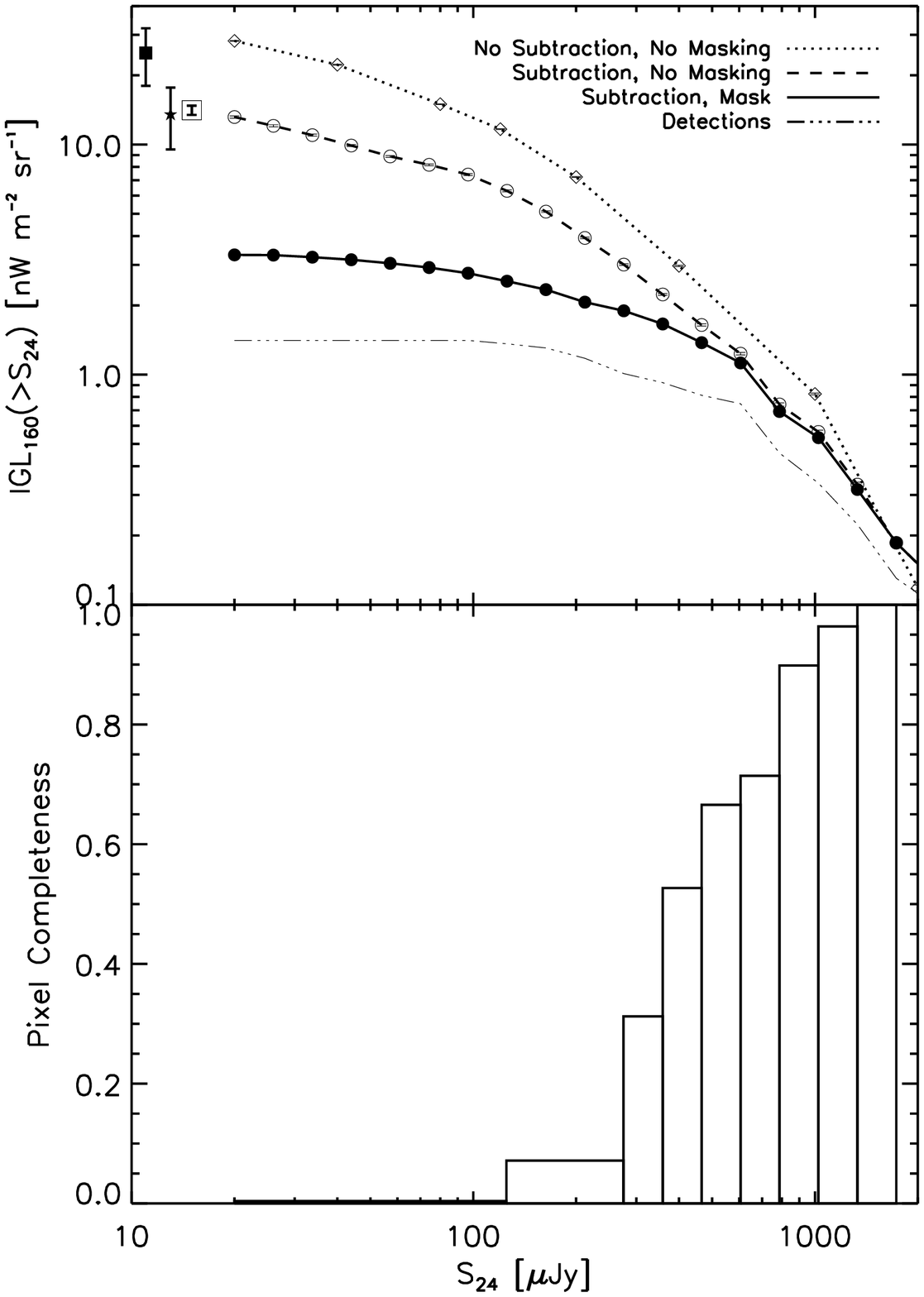}
\caption{Top Panel: The integrated galaxy
light at 70 (left) and 160\,$\mu$m (right) from sources brighter than a particular
24\,$\mu$m flux threshold, as estimated from {\it Spitzer} data
using different stacking methodologies. The solid black is the best estimate from the current data.
The EBL from DIRBE at 60 and 140\,$\mu$m is shown as the solid square, the EBL from FIRAS at 160\,$\mu$m as the solid star
while the model estimates presented in this paper are shown as the empty square. 
When bright sources that are individually detected are not subtracted and the stacking is done without masking out the pixels
that are included in the stack, the total IGL is significantly overestimated due to overcounting of the flux (dotted line). 
Even after the sources are subtracted, stacking without masking results in the stacked intensity 
being biased high. After masking,
large and uncertain completeness corrections are required to measure the IGL of sources below 
a 24\,$\mu$m flux density value of 74\,$\mu$Jy and 275\,$\mu$Jy in the 70 and 160\,$\mu$m images respectively.
This is shown in the lower panels as the fraction of pixels attributed to sources in differential 24\,$\mu$m
flux bins, that are unmasked and can contribute to the stacked flux.
Higher spatial resolution far-infrared data with {\it Herschel} will reduce the incompleteness due to masking and
improve stacked estimates of the IGL from faint 24\,$\mu$m sources.
The solid black line in the left panel shows the 70\,$\mu$m IGL after these completeness corrections are applied.
At 160\,$\mu$m, the completeness correction can only be applied above 275\,$\mu$Jy and results in values similar to the 
solid line. As a result, we omit it for clarity.
}
\label{fig:iglspitzer}
\end{figure}

We first subtract all reliably detected sources from the 70\,$\mu$m
mosaic. We do this by scaling the 70\,$\mu$m point spread function (PSF) to the flux density of each source
and subtracting it off.
The 127 reliably detected sources which have 70\,$\mu$m signal to noise ratios $>6$ and
which span a flux density range of $1.2-40.5$\,mJy are subtracted from the map. 
We conservatively choose 6$\sigma$ because the noise is correlated in the final mosaics as a result of which the measured noise
is biased low compared to the true noise. The exact magnitude of this effect is difficult to quantify due to the filtering
of the individual 70\,$\mu$m frames before coaddition, but we expect it to be a factor of $\sim$2.
These individually detected sources contribute
a total IGL intensity of 1.92$\pm$0.2\,\cirbu. We then take the residual map with these sources subtracted
and stack on the location of the remaining 1945 24\,$\mu$m sources spanning a flux density range
of $20-910$\,$\mu$Jy. We cut out a square region of width 84$\arcsec$ around the position of each source, starting
with the one with the highest 24\,$\mu$m flux density. 
We then mask out a region of radius 16$\arcsec$ around the nominal position of the source in the map
after including it in the stack. The cutouts are made in decreasing order of 24\,$\mu$m
flux density. The masked region is exactly the size of the aperture used to measure the photometry
and ensures that the flux in any particular pixel is not counted multiple times.
To stack, we take a simple sum of the pixel values from the cutouts although we note that taking a sigma clipped
sum does not result in any appreciable difference. Due to the mask, the image cutouts of
faint 24\,$\mu$m sources which are in the vicinity of brighter sources will mostly have values
of zero except for the unmasked pixels. However, since the pixel contribution from these faint sources
have been included when making the cutout for the brighter source, we ensure we are not double counting the flux
in any pixel. We obtain a strong detection in the resultant
stack with an IGL intensity of 1.7$\pm$0.7\,\cirbu. After application of a 70\,$\mu$m calibration
 correction factor of 1.15, this implies the total contribution
of the 24\,$\mu$m detected sources 
to the IGL should be 4.2$\pm$0.7\,\cirbu\ of which approximately half is from
the directly detected $>6\sigma$ 70\,$\mu$m sources (Figure \ref{fig:iglspitzer}).

We note that the masking 
is a critical step in the stacking process. In the absence of masking,
we would have counted the flux in the wings of brighter (but still undetected) sources and attributed this to 
faint sources. The IGL resulting from stacking the faint 24\,$\mu$m sources would then have been
3.2$\pm$1.1\,\cirbu, about
a factor of 2 bias in the contribution of the faint sources. After adding in the contribution of
detected sources and the 1.15 calibration correction factor, the resultant IGL would have 
been found to be 5.9$\pm$1.1\,\cirbu.

The quoted uncertainties in the stacked quantities are derived by performing Monte-Carlo simulations
whereby the positions of the 24\,$\mu$m sources are randomly offset by an amount which is between $1-3$\,FWHM.
This offset ensures that uncertainties in the stacked intensities due to imperfect subtraction of the 
brighter sources is taken into account. It also ensures that the intrinsic large scale clustering of 24\,$\mu$m 
sources is preserved in the estimation of uncertainties. Using a random set of positions distributed across
the entire map for the Monte-Carlo will underestimate the uncertainties due to these two effects.
The 70\,$\mu$m image is then stacked at these offset positions and combined by taking a simple sum.
The process is repeated a 100 times and the standard deviation of the aperture flux, with all appropriate aperture
corrections gives the uncertainty in the stacked value. 
%This technique ensures that the uncertainties due to the
%clustering properties of the sources and residuals from the fitting of the brighter sources
%are accounted for in the stacking uncertainty estimate.   
We find that these uncertainty values are typically 5
times higher than what one would derive by stacking on completely random positions on the source subtracted map.

The stacked value can change depending on the following effects. 
As expected, the total stacked flux is sensitive to the size of the masking region. 
If the masking radius and correspondingly, the aperture for photometry,
is increased to 35$\arcsec$, we find that the stacked flux decreases such that the total IGL, after adding the
contribution of detected sources and calibration corrections, is 3.3$\pm$0.7\,\cirbu.
The fraction of pixels that go into the stack is however reduced such that after completeness corrections, we
obtain a similar IGL to the one estimated with a smaller masking radius (see below).
Estimation and subtraction of a local background from each cutout before stacking
only results in a change of 1\% in the total IGL. Stacking on the 24\,$\mu$m positions without sorting
on 24\,$\mu$m flux density does not result in a significant detection.

The masking procedure however only includes a fraction of the pixels which correspond to faint
24\,$\mu$m flux sources contributing to the stack (Figure \ref{fig:iglspitzer}). 
For example, if all sources with 24\,$\mu$m flux densities above 25\,$\mu$Jy are considered,
only 37\% of the pixels of these sources are included in the stack while if the flux limit
is increased to 57\,$\mu$Jy, more than 54\% of the pixels of the sources are unmasked
and can be included. To correct for this, 
we need to measure the contribution to the stacked flux from sources in bins of 24\,$\mu$m flux and 
divide the stacked flux and uncertainty of each bin by the fractional number of pixels contributing to the stack. 
We call this a completeness correction\footnote{There is a second completeness correction associated with the
incompleteness of the 24\,$\mu$m catalog at the faintest flux densities. Since we dont know the locations of those
sources, we cannot include them into the simulations}.
The differential contribution to the IGL from 24\,$\mu$m
sources in a flux bin ($f_{1}<f_{2}$) can be calculated only by 
stacking on all sources brighter than $f_{1}$ in decreasing order of flux, 
including masking, and then subtracting the IGL derived from stacking on all sources brighter than $f_{2}$.
Simply considering sources between $f_{1}$ and $f_{2}$ and stacking on a mosaic with no sources subtracted
as has been undertaken by \citet{Dole} overestimates the flux although the exact overcounting factor depends
on the relatively uncertain clustering properties of the faint sources around bright sources.
We find that stacking on the 838 sources with S$_{24}>74\,\mu$Jy, after application of the
completeness corrections shown in Figure \ref{fig:iglspitzer} and appropriate calibration
corrections account for 4.9$\pm$0.5\,\cirbu.
The 1107
sources between $20-74$\,$\mu$Jy only have one fifth of their pixels included in the stack due to the masking.
The differential contribution to the stacked flux in this bin needs to be corrected upward by a factor of 5 yielding
2.1$\pm$0.9\,\cirbu. Thus, our best estimate for the IGL from the data, including detections, masking, completeness corrections and
calibration corrections is 7.0$\pm$1.0\,\cirbu. We note that the 24\,$\mu$m catalog itself
is 85\% complete in the faintest flux bins
but we neglect application of this second order completeness correction which would add 0.4\,\cirbu\ to the best IGL estimate
derived above.

An alternate technique that we adopt
to measure any bias in our stacking technique is by simulating the data. We input point
sources into a simulated 70\,$\mu$m map at the locations of the 24\,$\mu$m sources that we stack on. 
The 70\,$\mu$m/24\,$\mu$m flux ratios for these simulated sources are derived from the CE01 model templates and the
evolution of the infrared luminosity function derived by \citet{Magnelli}. 
We also make sure that this map does not contain any $6\sigma$ sources so we are trying to simulate our residual map which 
we use for stacking. 
We then
repeat the stacking procedure on this simulated image
using no masking and compared the stacked flux with the total input flux.
We find that the stacked flux exceeds the input flux by a factor of 1.2. Thus, using the simulations, we estimate
that the IGL in the maps is probably the sum of 1.92 (detected sources) and 3.2/1.2, which after calibration correction
factors implies 5.3$\pm$1.1\,\cirbu although the exact value may be susceptible to the redshift distribution of the
sources and the distribution of their 70\,$\mu$m/24\,$\mu$m flux ratios.
 
Our analysis therefore suggests that the true 70\,$\mu$m IGL from 24\,$\mu$m galaxies brighter than 25\,$\mu$Jy is in the range
$5.3-7.0$\,\cirbu, about 30\% of which is due to galaxies that are individually detected in the 70\,$\mu$m map.

A comparison with previous estimates of the 70\,$\mu$m IGL reveals that \citet{Dole} estimated
an IGL of 5.93$\pm$1.02\,\cirbu\ for source with
S$_{24}>60\mu$Jy. By adopting a total EBL estimate of 6.4\,\cirbu, they claim
that the bulk of the IGL is resolved. We next attempt to reproduce these measurements using an identical set of steps
as described in the \citet{Dole} paper. We divide the 24\,$\mu$m source list into flux bins and stack on the sources
in each bin separately, with no masking. The stacking is performed on the maps which still
have the individually detected sources in them as was done by \citet{Dole}. 
We then add the contribution of the stacked measurement in each 24\,$\mu$m flux bin and
obtain a 70\,$\mu$m IGL value of
7.1$\pm$1.4\,\cirbu\ for S$_{24}>60\mu$Jy (Figure \ref{fig:iglspitzer}). This is a significant overestimate since the stacking on the sources
in the fainter flux bins includes sources which fall within the PSF of brighter sources. As a result, the flux in those
pixels is counted multiple times. 
If we 
mask the sources as described earlier, we derive a value of 3.3$\pm$1.0\,\cirbu for sources with S$_{24}>60\mu$Jy which is significantly
below the \citet{Dole} estimates. However, after completeness for the fraction of pixels missed
and calibration corrections, this number rises to 5.4$\pm$0.5\,\cirbu, as discussed 
above. Although the difference in the derived value compared to the \citet{Dole} estimate
is not large, it is clear that a systematic bias is introduced
depending on how the stacking is done. As we will demonstrate in the subsequent sections, this
systematic bias gets larger with a larger point spread function. Thus, we believe that each of the steps: source subtraction, 
stacking on the residual maps with masking and simulations of the bias need to be performed to obtain a reliable stacked result.

\citet{Frayer} estimated the IGL by integrating
over the 70\,$\mu$m source counts distribution in GOODS-N, without stacking. They found 
that the contribution of directly resolved
sources with S$_{70}>1.2$\,mJy is 4.3$\pm$0.7\,\cirbu. This is consistent with our values derived from
stacking since it is simply an integral over a source counts distribution.
\citet{Frayer} derive a total EBL of 7.4$\pm$1.9\,\cirbu by adopting a power-law
for the differential counts and extrapolating
the source counts to zero flux. 

The luminosity function of \citet{Magnelli}, if evolved between $0<z<1.3$, predicts a 70\,$\mu$m IGL
from $z<1.3$ galaxies of 9.2\,\cirbu. This is using the CE01 far-infrared templates which has been
shown by \citet{Magnelli} to be a good fit to the 24$\mu$m/(1+z) and 70$\mu$m/(1+z) luminosities of galaxies.
Thus, we conclude that the total EBL estimate of \citet{Frayer} is a significant underestimate.
As we will discuss later, evolutionary models predict that the fraction of the 70\,$\mu$m EBL which can be
attributed to galaxies with $S_{24}>25\mu$Jy is $\sim$80\% with a total EBL of 9.5$\pm$0.3\,\cirbu. 
Since our measured IGL from stacking is between $5.3-7.0$\,\cirbu, the deepest current observations at 70\,$\mu$m 
have resolved about $\sim$56-75\% of the EBL.

\subsubsection{160\,$\mu$m}

At 160\,$\mu$m, we utilize the publicly released first epoch FIDEL data in the Extended Groth Strip.
We analyze only the deepest 914\,arcmin$^{2}$ which has a total exposure time of $>$500s at 160\,$\mu$m. The 
corresponding coverage in the 24\,$\mu$m map is $>$7200s. 
Details of the 160\,$\mu$m mosaic are in Table \ref{tab:blast}.
Starting with the 3.6\,$\mu$m catalog,
we fit for the fluxes of the 24\,$\mu$m sources using the same technique as above and which is
described in \citet{Magnelli}.
We then apply a flux density cut of 100\,$\mu$Jy to the 24\,$\mu$m catalog. The coordinates of these
 brighter 24\,$\mu$m sources
are provided as input to the point source fitting in the 160\,$\mu$m image. The uncertainty in the
160\,$\mu$m photometry of each source
is estimated as the sum of the squares of the residuals in the 160\,$\mu$m map
after all sources are fit and subtracted. We apply a signal to noise
cut of $>$5$\sigma$ and 20\,mJy in the 160\,$\mu$m catalog for reliability.
We find that these reliably detected 160\,$\mu$m sources contribute 
1.4$\pm$0.1\,\cirbu. After subtracting out these sources, we stack at the positions
of the remaining 24\,$\mu$m sources following the guidelines described earlier for the 70\,$\mu$m stacking.
The public release of the 24$\,\mu$m EGS observations are relatively shallow compared to the GOODS-N data
described earlier, but still deeper than the {\it Spitzer} observations used by \citet{Dole}.
We estimate the 24\,$\mu$m data in EGS is complete at 55\,$\mu$Jy depth.

Stacking on sources detected with high confidence in the 24\,$\mu$m map, including masking, results in a stacked
160\,$\mu$m intensity of 1.9$\pm$0.3\,\cirbu implying a total 160\,$\mu$m IGL intensity of
3.3$\pm$0.4 \cirbu. We note that the masking radius used (Table \ref{tab:blast}) is 
large compared to the FWHM of the PSF due to the fact that the PSF in the broad 160\,$\mu$m
filter is more sensitive to the intrinsic spectrum of the source in the sense that red sources
have broader PSFs than blue sources. To avoid missing flux, we use a larger aperture and correspondingly
larger masking radius.
If we had chosen not to mask the sources after including them in the stack, we
would have derived a total stacked intensity of 11.8$\pm$1.6\,\cirbu and a total IGL of 
13.2$\pm$1.7\,\cirbu after including the contribution of detected sources (Figure \ref{fig:iglspitzer}).

As we did for the 70\,$\mu$m data, we need to apply completeness corrections for the 
pixels of the stacked sources that have been masked due to their presence near a brighter 24\,$\mu$m source. 
The fractional incompleteness in the 160\,$\mu$m stacking at the same 24\,$\mu$m
flux density is higher than at 70\,$\mu$m since the point spread function is larger (Figure \ref{fig:iglspitzer}).
We find that the fraction of pixels included in the stack is
39\% if all sources brighter than S$_{24}>212\mu$Jy are stacked together.
This number increases to 52\% at 275\,$\mu$Jy. Thus, after completeness corrections to the stacking analysis
from sources with S$_{24}>275\mu$Jy and including the contribution of individually detected
sources, we can account for 2.9$\pm$0.3\,\cirbu of the 160\,$\mu$m IGL. We cannot 
account for the contributions of fainter 24\,$\mu$m sources between $20-275$\,$\mu$Jy since only 1.7\% of the pixels
from these sources are included in the stack. The resultant completeness corrections are large and unreliable although
the stack from this small fraction of sources that can be stacked on does result in a significant detection.

We can however estimate a lower limit for their contribution by stacking down to the maximum flux limit $f_{1}$ such that
the contribution of sources between $f_{1}$ and 275\,$\mu$Jy results in a significant stacked detection. We find that 
in the 24\,$\mu$m flux density range of 125$-$275\,$\mu$Jy, we can get a measurement of the stacked flux
although with a completeness correction factor of
$\sim$14. The resultant contribution to the IGL from sources in this bin, after completeness corrections,
is 4.2$\pm$0.9\,\cirbu. Thus, based on the stacking analysis
as much as 7.1$\pm$1.0\,\cirbu of the 160\,$\mu$m IGL may be resolved. 

As we did for the 70\,$\mu$m data, we attempt to simulate the 160\,$\mu$m map to assess how large the bias is
if we stack on the residual map with no masking. 
We input point
sources using the 160$\,\mu$m PSF at the locations of the 24\,$\mu$m sources that we stack on.
The 160\,$\mu$m/24\,$\mu$m flux ratios for these simulated sources are derived from the CE01 model templates and the
evolution of the infrared luminosity function derived by \citet{Magnelli}. We then
repeat the stacking procedure on this simulated image 
using no masking and compared the stacked flux with the total input flux.
We find that the stacked flux exceeds the input flux by a factor of 1.4. Thus, using the simulations, we estimate
that the IGL in the maps is probably the sum of 1.4 (detected sources) and 11.8/1.4, which 
implies 9.8$\pm$2.0\,\cirbu although the exact value may be susceptible to the redshift distribution of the
sources and the distribution of their 160\,$\mu$m/24\,$\mu$m flux ratios.

We conclude that the measured 160\,$\mu$m IGL derived from the deepest
{\it Spitzer} maps is 3.3\,\cirbu\ although simulations and estimates
of completeness at the faint flux bins yield a value between 7.1$-$9.8 \cirbu. In comparison, \citet{Dole} estimate an IGL
of 10.7$\pm$2.3 by stacking on sources with S$_{24}>60$\,$\mu$Jy
from their shallower imaging data. We note that if we follow the \citet{Dole} stacking prescription in this field,
we obtain a contribution of 15.0$\pm$2.0\,\cirbu from S$_{24}>60$\,$\mu$Jy sources. The 
overestimation in the stacking derived IGL at 160\,$\mu$m
due to the overcounting of flux in individual pixels is illustrated in the plots
shown in Figure \ref{fig:iglspitzer}. 

We also note that due to the large point spread function, it is not possible to reliably measure
the contribution to the 160\,$\mu$m EBL from sources below S$_{24}<275\,\mu$Jy. At this flux level, less
than 50\% of the pixels from faint 24\,$\mu$m sources contribute to the stacked flux and the differential
contribution from sources between $120-275$\,$\mu$Jy cannot be measured to better than 5$\sigma$. Higher resolution
data with {\it Herschel} will enable the contribution of fainter sources to be measured as described later.

The total EBL at 160\,$\mu$m derived by \citet{Fixsen} using FIRAS data is 13.6$\pm$0.4 \cirbu
while our models described later yield a value of 14.1$\pm$0.7\,\cirbu. Thus, the current observations have not 
resolved the 160\,$\mu$m background. 

\subsection{BLAST Submillimeter Survey}

BLAST is a 1.8m submillimeter telescope on a balloon, with detectors at three 
wavelengths: 250, 350 and 500$\,\mu$m (Pascale et al.~2008). BLAST completed a 11 day 
flight from Antarctica in December 2006 during which it took very 
deep images of the Extended Chandra Deep Field South (Devlin et al.~2009).

The BLAST data were released to the public in April 2009\footnote{http://blastexperiment.info/results.php}.
This release includes several versions of the signal and noise maps at all three wavelengths. The ``decon.fits" files (hereafter referred to as the decon maps) have had a whitening filter applied to suppress the large scale frequencies -- this affects the PSF and causes it to have negative structure and zero mean. The ``smooth.fits" files have been convolved with the PSF in order to estimate the point source fluxes. Convolving with the PSF is equivalent to a minimum $\chi^{2}$ fitting of the PSF with the data (see Serjeant et al. 2003). This method is generally used to extract point sources in submm surveys as it increases the signal to noise ratio (SNR) and aids detection of faint ($>3\sigma$) sources.
However, this technique works under the assumption that a single PSF is a good fit to the data and may not be valid when there are blended and confused sources (it is only valid if there is $<<1$ source per beam, Serjeant et al.~2003).

Point source catalogs down to $3\sigma$ from these BLAST maps are given in Devlin et al.~(2009).
Information about the PSF and pixel scale for the BLAST maps is given in Table \ref{tab:blast}.

\subsubsection{250, 350 and 500\,$\mu$m}

Unlike in the infrared, where stacking is done by excising cutouts of the image, coadding the cutouts
and doing aperture photometry, in the submillimeter,  stacking is usually done on the 
smoothed (beam-convolved) maps where single pixel values, which contain the total flux of each source,
are averaged to produce the stacked 
average \citep[e.g.][]{Devlin}. This is primarily because of the complex beam shape 
which can have significant sidelobes. The sidelobes can contaminate the aperture photometry of other
sources, resulting in a systematic uncertainty to the aperture flux which is dependent
on the flux distribution and separation of the sources (See Appendix for details). 

Although stacking on the BLAST maps has been presented in three separate 
papers \citep{Devlin, Marsden, Pascale}, there are certain improvements
which can be made to the analysis presented therein. 
In Marsden et al.~(2009) it is argued that stacking essentially boils down to taking the covariance of the map with the catalog. This simplification is true under the assumption that galaxies are not clustered. We know that IR luminous galaxies are clustered 
more strongly than optically selected field galaxies
\citep{Gilli}. Although the clustering of IR luminous galaxies has been measured in redshift bins, it is
the clustering strength of faint IR sources in the vicinity of brigher IR sources that 
is difficult to measure. We find that the surface density of 24\,$\mu$m detected sources within a radius R (Table \ref{tab:blast}) of
a 500\,$\mu$m source brighter than 20 mJy is twice as high as the average surface density of 24\,$\mu$m detected sources
across the entire field (Figure \ref{fig:cluster24} in the Appendix).
Results from stacking are systematically biased high by this clustering.
As a result, the stacking method employed by \citet{Devlin, Marsden, Pascale} will result in values
biased high since the flux from the many clustered sources within each BLAST beam will essentially be counted more than once. 
Serjeant et al.~(2008) estimated this over-counting of the stacked flux assuming a clustering strength and showed that it was significant even at the much smaller beam of SCUBA on the JCMT. 
There are severals steps that can be taken to test and account for this effect of clustering on the stacked signal which we apply to our own BLAST stacking analysis below.

%As a result, we revisit the BLAST stacking analysis.
We perform our BLAST stacking analysis on both the decon maps where we stack cutouts and the smoothed maps where we stack single pixel values. We find consistent results between these
two approachs. Values quoted in this paper are all for stacking image cutouts on the decon maps.

As described earlier
for the 70\,$\mu$m and 160\,$\mu$m stacking analysis, we start off with the list of 24\,$\mu$m sources
cataloged for the ECDFS by \citet{Magnelli}.
%The area of the BLAST PSF at 250\,$\mu$m 
%within the FWHM is 0.2 arcmin$^{2}$ which means that on average there are 
%several 24\,$\mu$m sources within each BLAST beam. Furthermore, 24\,$\mu$m sources are clustered
%relatively strongly \citep[e.g][]{Gilli} implying that the number of sources within the beam might
%be even higher than what is expected from Poisson statistics.
Building on the techniques presented for the {\it Spitzer} far-infrared data,
we take three additional steps which many previous stacking analysis in the submillimeter have not accounted for:
\begin{itemize}
\item We fit the bright detected submillimeter sources and subtract their flux from the map. This residual map will then be used for the stacking analysis. The bright point sources leave a substantial footprint on the map which causes stacked fluxes to be overestimated if they are left in the map. The contribution to the background from the detections will be counted separately after doing the stacking analysis.

\item We mask regions of the map as they are included in the stack. Again the large beam means that there are many 24\,$\mu$m sources per beam. If we don't exclude regions of the map that are already in the stack then flux 
in those regions will be counted more than once.

\item We perform simulations of confused maps with different source count distributions to assess the presence of
any bias in the stacked intensity values.
\end{itemize}

We calculate the uncertainties on all stacked quantities by doing Monte Carlo simulations stacking 
random positions in the map. We performed 10000 trials of stacking $N$ galaxies (where $N$ 
is the number of galaxies in the catalog). The distribution of stacked values from the 10000 
trials is confirmed to be Gaussian centered around zero and we take the standard deviation of 
the distribution as our uncertainty in the stacked flux.

\subsubsubsection{Subtracting the submillimeter detections}

\begin{figure*}
    %\centering
    %\includegraphics[width=6.0in]{f2.eps}
    %\vspace{0.1in}
    \plotone{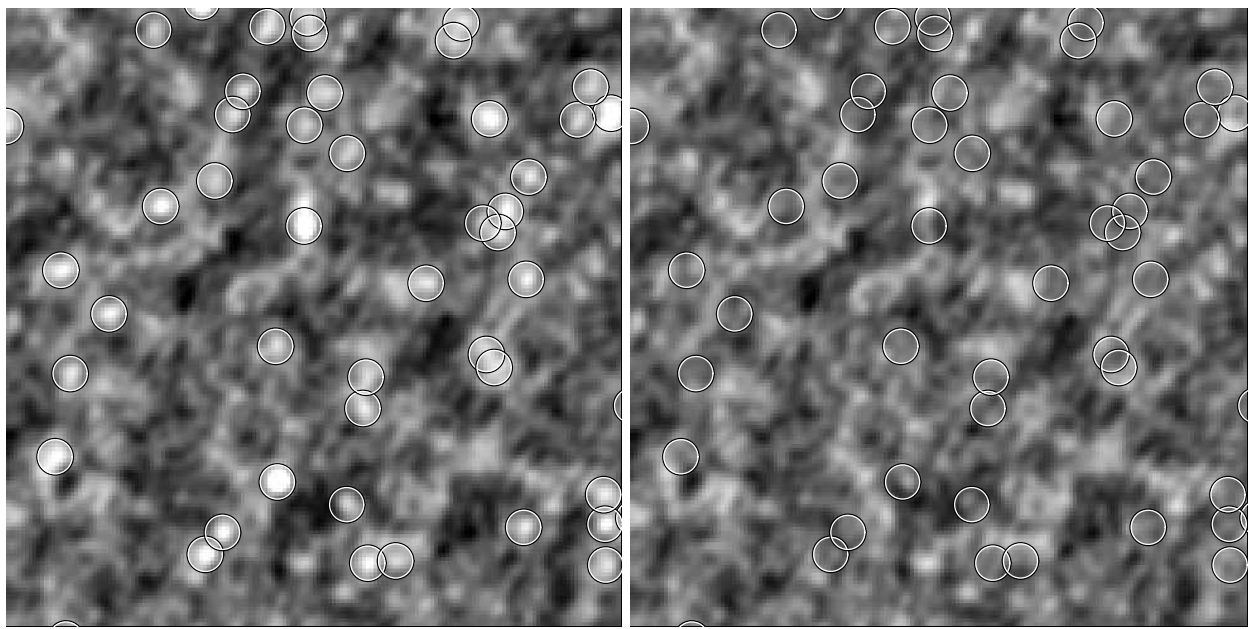}
    \caption{Subtracting the submillimeter detections from the maps prior to stacking. Images are 27$\arcmin$ across. Circles have a radius of 45$\arcsec$ and indicate the positions of the $5\sigma$ detections. The left panel shows the raw (decon) 250$\,\mu$m map and the right panel is the same map with the detections removed. }
    \label{fig:clean}
\end{figure*}

In order to test the effects of bright sources on the stacked signals, we iteratively fit the sources out of the map. 
%This technique was successfully applied to the SCUBA 850 $\mu$m map of GOODS-N \citep{Pope05}. 
Starting with the brightest 250 $\mu$m source, we fit the actual PSF (including positive and negative structure)
to the decon image at the source position and remove the signal from the decon map. We continue to the next brightest source and so on. 
For the positions of the detected sources we use the source lists from Devlin et al.~(2009) down to $4\sigma$.
After removing all the 250 $\mu$m emission from detected sources above a given SNR threshold, we are left with a residual map (see Figure \ref{fig:clean}). The distribution of pixel values in this residual map compared to the input map is shown in Figure \ref{fig:cleandist} in the Appendix.
We stack this map on the 24 $\mu$m galaxies. 

We calculate the total flux removed from the maps from the detected sources. Table \ref{tab:detected} lists the contribution to the background from sources detected above 6, 5 and 4$\sigma$ in the central region of the BLAST maps that overlaps the deepest 24 $\mu$m coverage.

We note that there is no bias associated with
estimating the contribution from the detected sources separately from the stacked sources based on the covariance argument in Marsden et al. (2009). The only complication associated with the subtraction of sources from the maps might be flux boosting but since source
subtraction is limited to the highest SNR sources and we are fitting the full PSF (and not simply scaling to the peak flux), flux boosting is expected to be much less of an issue. 

The results of stacking on the residual maps are shown in Table \ref{tab:blast_stack} and 
Figures \ref{fig:250} (bar plots). Leaving 
the detected sources in the maps prior to stacking, results in a noticeable 
over estimate of the stacked flux as was shown for the 70 and 160\,$\mu$m data. 
By simply subtracting the submillimeter detections prior to stacking we found that the IGL resolved by 24$\,\mu$m-selected galaxies is $\lesssim50\%$ the total EBL at these submillimeter wavelengths.
Stacking on the residual maps will still result in an overestimate since the flux surrounding sources $<4\sigma$ can still be double counted. We can't subtract these sources individually but we can mask them so their flux is not counted multiple times. 

\subsubsubsection{Masking while stacking}

We stack image cutouts from the BLAST
maps centered on each 24 $\mu$m source which is not matched to a detected submillimeter source. Once a cutout is made from the map and added to the stack, these pixels are masked out in the original map so that the same flux cannot be counted more than once. 
We step through the 24 $\mu$m catalog in order of decreasing 24 $\mu$m flux since the brighter 24$\mu$m sources are expected to have higher 250$\mu$m flux on average.
We ensure that the mean of the map is zero prior to beginning our stacking analysis which means that we do not need to do any additional sky-subtraction from our stack.

We tried a range of cutout sizes and settled on the values given in Table \ref{tab:blast} since those require a small aperture correction and doesn't mask out too much additional flux. Assuming a perfect Gaussian beam with 
FWHM of 36$\arcsec$ at 250$\,\mu$m \citep{Marsden} this cutout size should include all flux within $\pm3$ times the standard deviation
of the Gaussian (in reality the beam is not a perfect Gaussian).

We measure the stacked flux from the cutout by fitting the actual PSF. This is better than fitting a simple Gaussian or doing aperture photometry since the negative sidelobes in the PSF is significant. We tested this on cutouts of detected 250\,$\mu$m sources and found that it gave better results than aperture photometry or fitting a Gaussian (see Figure \ref{fig:apcorr}). From this test, we derive a negligible aperture correction of 1.02 to account for the emission outside the cutout area. 

\begin{figure}
    %\centering
    %\includegraphics[width=3.0in]{f3.eps}
    %    \vspace{0.1in}
    \plotone{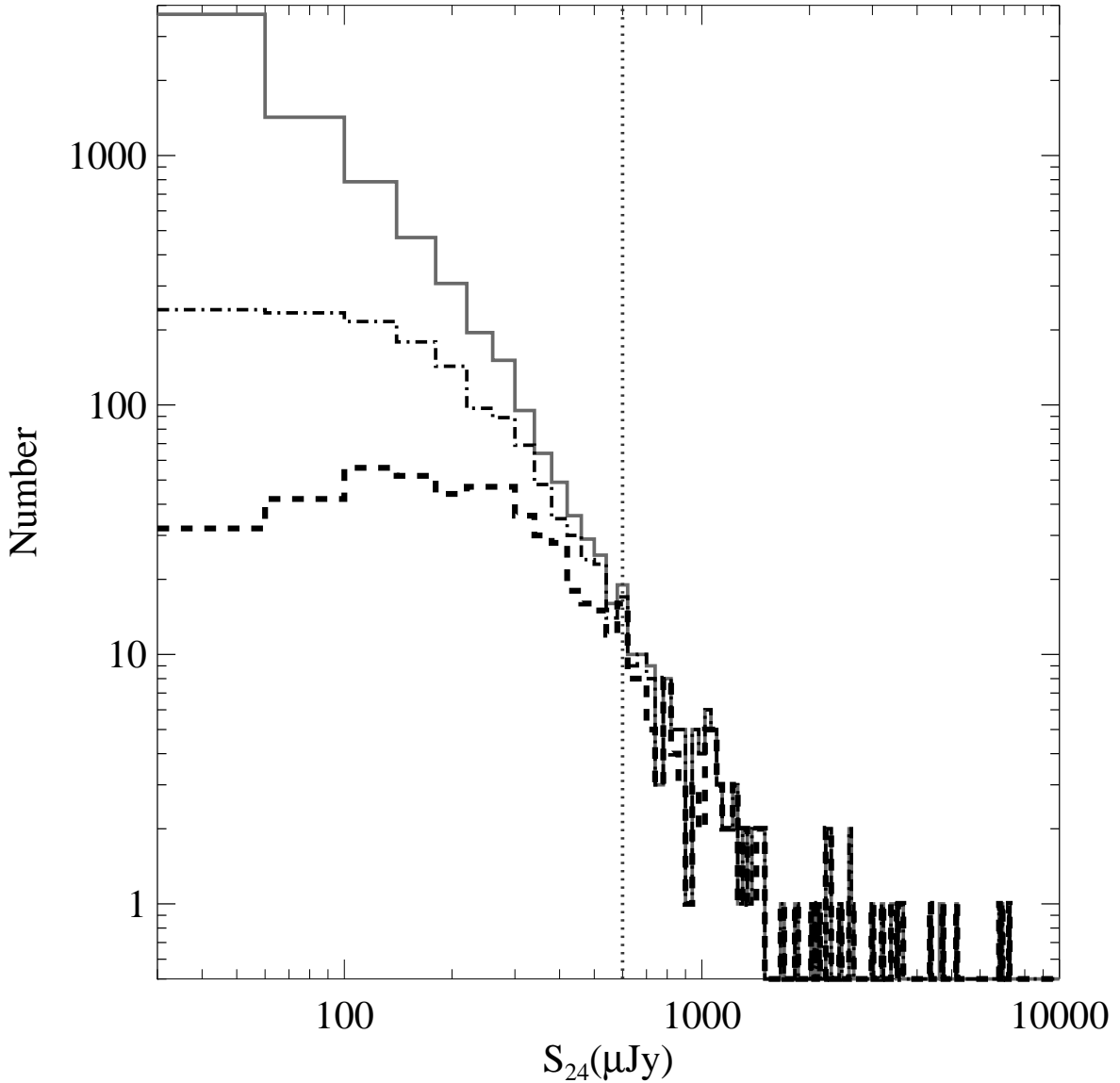}
    \caption{Distribution of 24 $\mu$m flux densities of sources that can be stacked on at 250\,$\mu$m after masking.
The grey solid curve shows all secure 24 $\mu$m detections in the ECDF-S and the black dashed curve is the flux distribution for the sources we are able to stack in the 250 $\mu$m map with masking. The vertical dotted line shows the flux at which we are no longer complete in our stacking with masking. With only a few tens of sources we do not obtain a detection when stacking differential bins below $S_{24}=600$\,$\mu$Jy.
The dot-dash line is the flux distribution of sources from stacking with masking on a map with same spatial resolution as 250$\,\mu$m with {\it Herschel}; with 100s of sources per bin {\it Herschel} will be able to constrain the contribution to the IGL from sources with $S_{24}=20$--$600$\,$\mu$Jy.}
    \label{simulationfigure}
\end{figure}

\begin{figure*}
\begin{center}
\includegraphics[width=3.0in,angle=0]{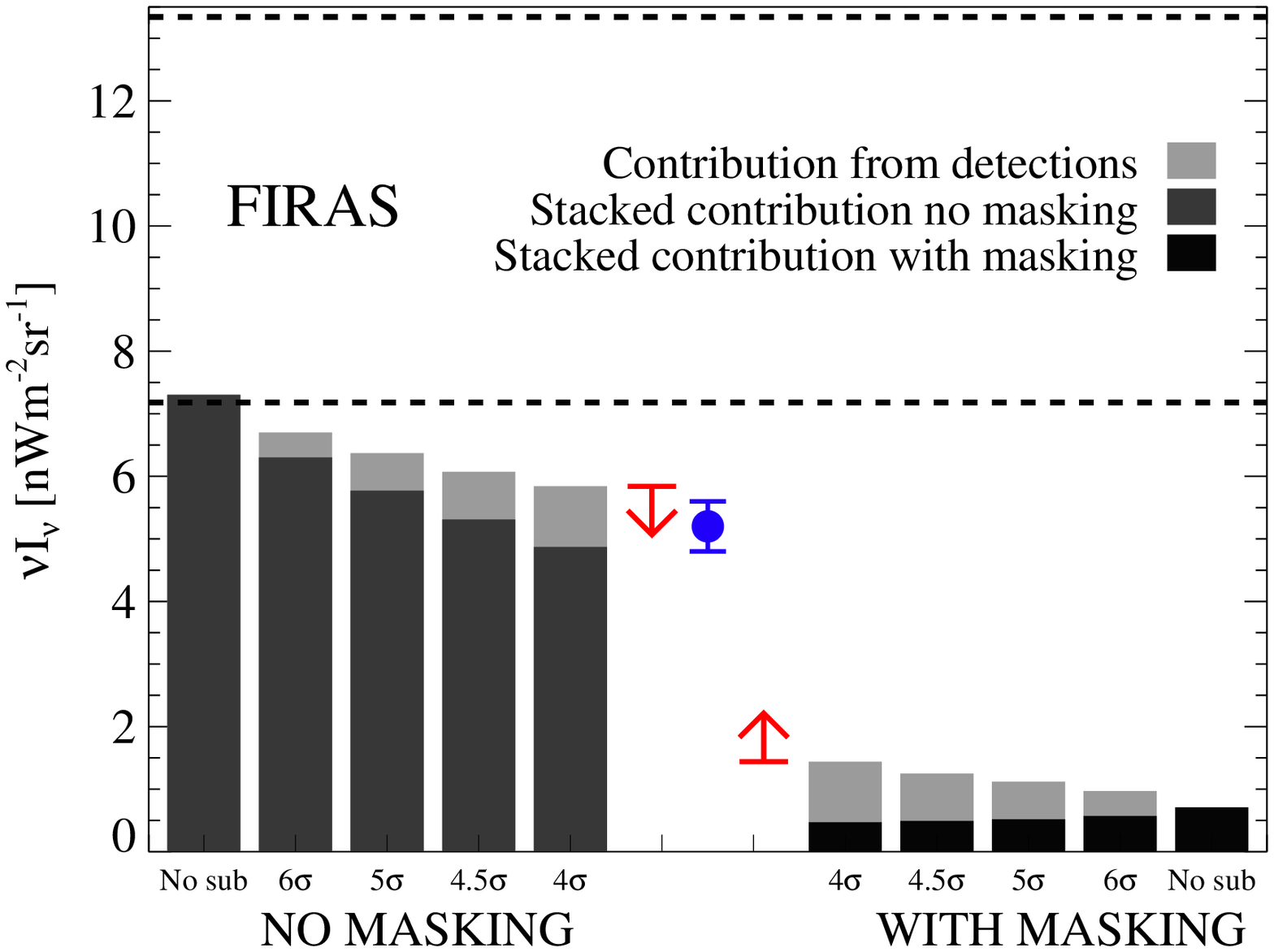}
\includegraphics[width=3.0in,angle=0]{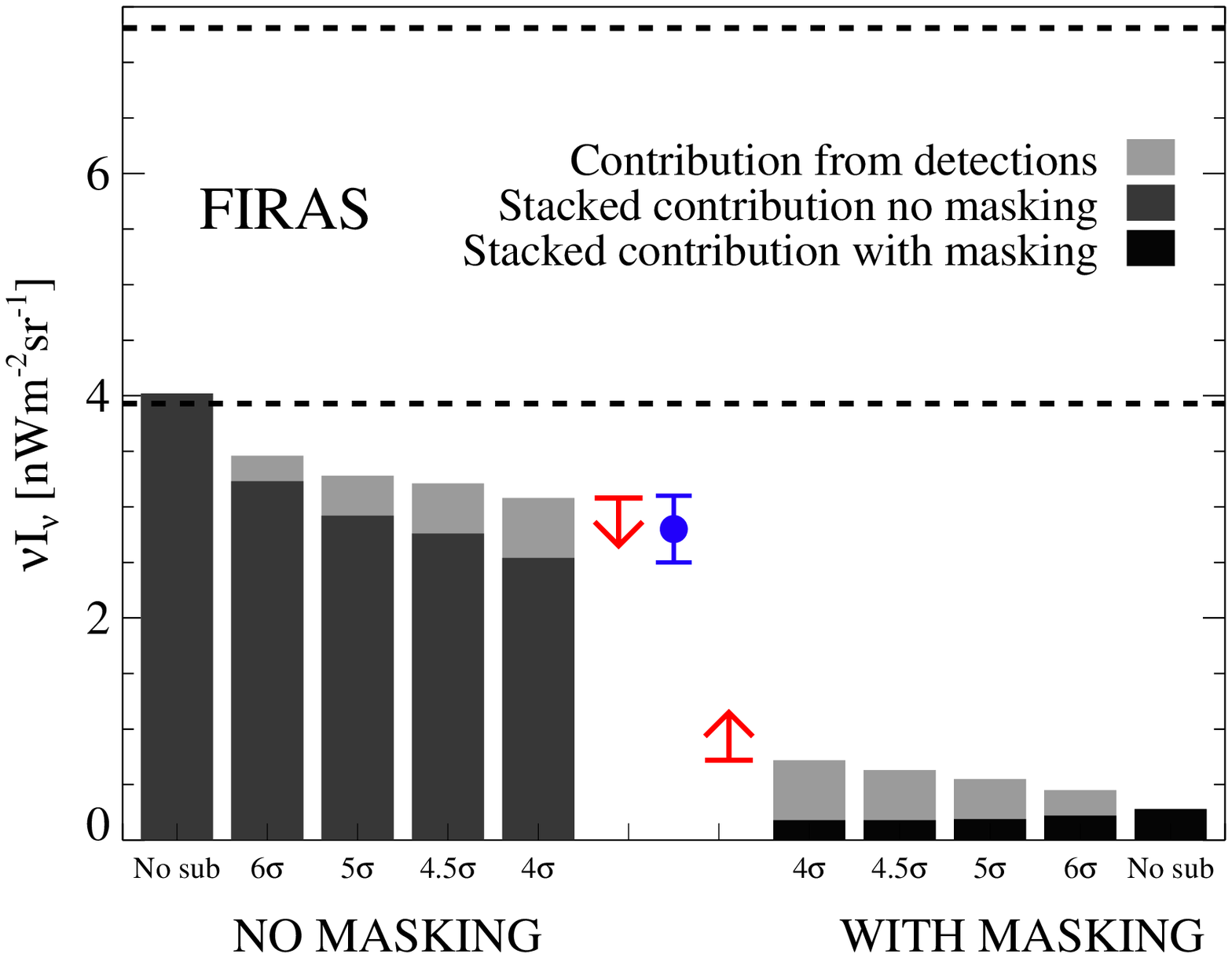}
\includegraphics[width=3.0in,angle=0]{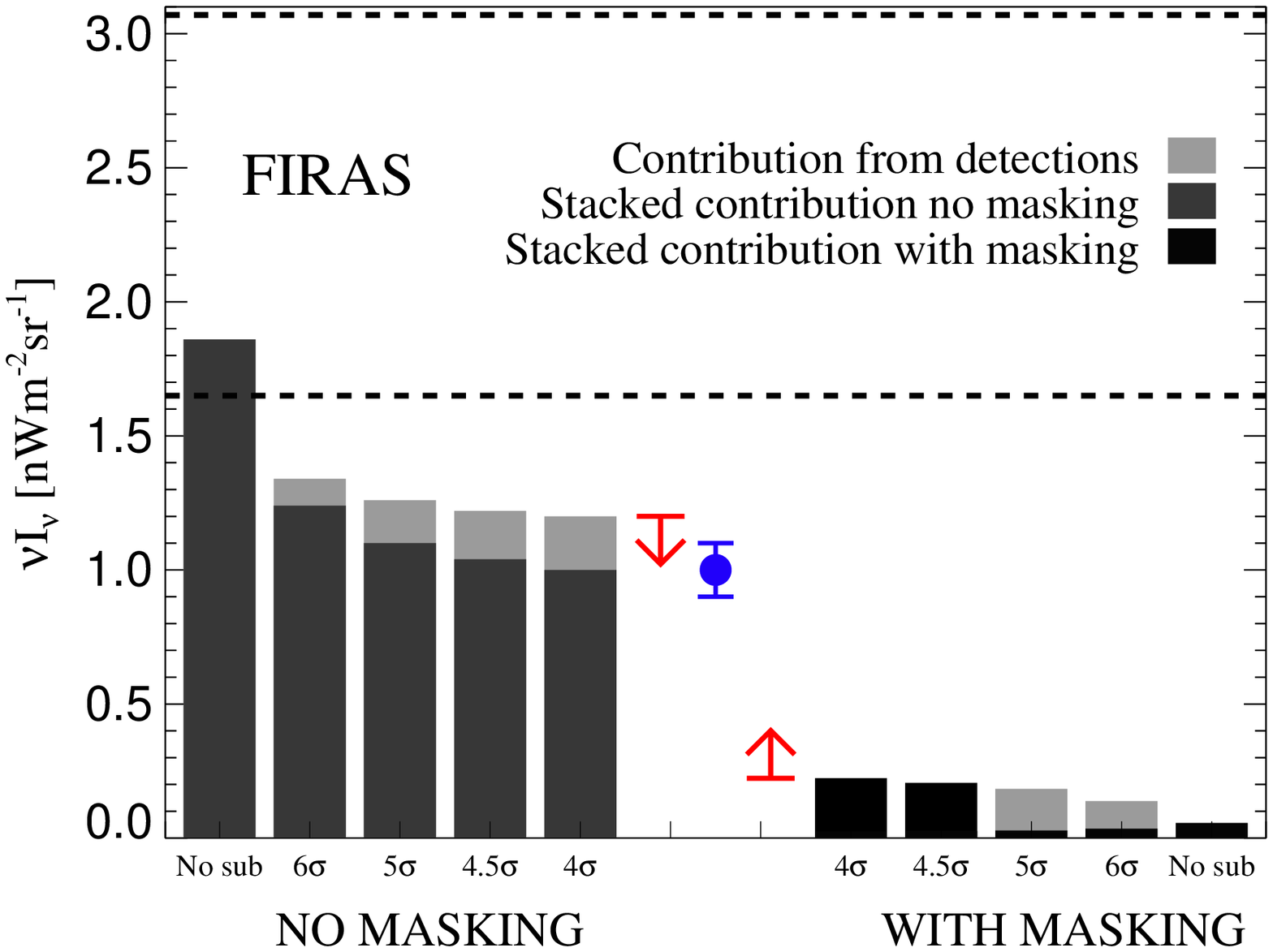}
\caption{Contribution of 24$\,\mu$m galaxies to the 250 (top left), 350 (top right) and 500$\,\mu$m (bottom)
submillimeter background. Light bars are the contribution of bright, individually detected submillimeter sources
while dark bars are the contribution from stacking on 24\,$\mu$m sources that are not associated with the bright submillimeter 
sources. The bottom axis indicated the signal-to-noise
threshold down to which detected sources were subtracted from the maps. The bars on the left 
are for stacking without masking -- these measurements will be biased high (as shown by the lower limit). The bars 
on the right are for stacking with masking -- these measurements will be biased low (as shown by the lower limit). The blue data point is 
our best estimate for the background from all 24$\,\mu$m sources ($S_{24}>40\,\mu$Jy) derived by simulating the stacking experiment.
}
\label{fig:250}
\end{center}
\end{figure*}

When masking while doing the stack, we are only able to stack a small fraction of the 24$\,\mu$m sources. 
Since we are stacking in order of decreasing 24$\mu$m flux we can find the flux at which there is no regions of the map left to stack. The 24$\mu$m flux below which we begin to be incomplete is 600$\mu$Jy (see Figure \ref{simulationfigure}). We note that this flux limit is comparable to that obtained
at 160\,$\mu$m (see Fig.~\ref{fig:iglspitzer}) which is consistent since the beams are almost the same size. 
We also note that this is the flux limit where there are $\sim$6 250\,$\mu$m BLAST beams/source which is well
past the standard confusion limit of 40 beams/source and even past the 7 beams/source
limit where prior based source extraction techniques start to give inadequate results \citep{Magnelli}.

As we did for the 70 and 160\,$\mu$m analysis,
we explored applying completeness corrections to the bins below 600\,$\mu$Jy. The stacked fluxes for the individual bins are
however not significant
and so we can't reliably apply a correction to account for the sources we have had to mask. In a very wide area survey, where 
there are 100s of objects in each bin below 600\,$\mu$Jy then we could imagine applying a completeness correction to 
the masked stacking values to resolve more of the background. This will be possibly with {\it Herschel} where the beam is half the size so many more sources can be stacked with masking. 

Our results from stacking with masking are shown in Table \ref{tab:blast_stack} and 
Figures \ref{fig:250}. Since we are unable to apply completeness corrections, these values are lower limits to the true IGL from the 24$\,\mu$m galaxies; the true value will lie somewhere in between the masking values and the no masking values. 
In addition, we expect the stacking with masking values will be biased low since we have only masked the positive emission in the zero-mean BLAST PSF and therefore are stacking on a map with a slight negative mean.
From our masking analysis, we conclude that with
the spatial resolution of the current data, the lower limit to the IGL from 24$\,\mu$m selected galaxies between $250-500$\,$\mu$m
is $\sim$10-15\% of the total EBL shown in Table 5.

\subsubsubsection{Simulations}

We can further test how the stacking analysis might be biased by performing simulations where we know 
the IGL we put into the map (IGL$_{in}$) and then we perform the same stacking procedure 
to see what we measure (IGL$_{out}$). The ratio of these two quantities is then our over-counting or 
bias factor. This can give us an estimate of the true IGL coming from the 24$\,\mu$m-selected galaxies. 

We performed simulations of the BLAST 250$\,\mu$m maps by doing the following (the same steps were repeated for the 350 and 500$\,\mu$m maps as well): 

1) Assign a 250$\mu$m flux to each 24 $\mu$m source using the full range of conversions from the CE01 galaxy templates. These sources are inserted into the fake map using the actual BLAST PSF at the real 24$\,\mu$m positions so as to include the inherent clustering of this population in our simulation.  

2) Add noise to the fake map according to the actual BLAST noise maps.

3) Subtracted any $4\sigma$ detections from the maps to create a fake residual map for stacking. 

4) Perform stacking at positions of all input galaxies to get IGL$_{out, no mask}$.

5) The ratio of IGL$_{in}$/IGL$_{out, no mask}$ tells us how much we are over-predicting the stacked flux when we stack on the residual maps without masking. This bias factor is listed in Table \ref{tab:over}.

6) We can also do the stacking with the masking to get IGL$_{out, mask}$ and the ratio IGL$_{in}$/IGL$_{out, mask}$ tell us how much we are under-predicting the stacking flux when we mask.  

7) Iteratively determine the IGL$_{in}$ which reproduces both our observed IGL$_{out, no mask}$ and our observed IGL$_{IN, mask}$. This value is referred to as IGL$_{corr}$.

The value of IGL$_{corr}$ is an estimate of
the actual background produced from the 24 $\mu$m galaxies (assuming all 250 $\mu$m detections have 24 $\mu$m counterparts) 
and is plotted as the blue circle in Figures \ref{fig:250} and given in Table \ref{tab:over}.

We note that our values are significantly below the FIRAS determined EBL values as measured
by \citet{Fixsen} and therefore conclude
that $\sim$45-55\% of the submillimeter background currently remains unresolved. This conclusion is different from that of \citet{Devlin} and \citet{Marsden} who do not do any masking or subtract the submillimeter detections prior to stacking. We have demonstrated that these two steps are crucial to ensure that the results are not biased due to overcounting the flux from multiple sources within the large beam because of the clustering of IR luminous galaxies. 
Aside from differing stacking techniques we also show later in this paper that we do not expect 24$\,\mu$m-selected galaxies to resolve all the EBL since there are significant amounts of star formation coming from less luminous galaxies.
We also note that the ``completeness" corrections that \citet{Devlin} apply to the FIDEL 24\,$\mu$m catalogs are small but
incorrect. The completeness
corrections they apply are based on much shallower data than the FIDEL data and the prior based cataloging technique used to generate
the FIDEL 24\,$\mu$m catalogs are less affected by confusion, resulting in higher completeness.

This bias factor listed in Table \ref{tab:over} provide a rough guide as to how much the stacked signal will overestimate
the true value, if one does not correct for the contribution of 
multiple sources per beam due to the clustering of IR luminous galaxies. If all things were equal then this bias factor would only depend on the FWHM of the beam and 
the density of sources that are being stacking. However, in practice the bias factor will also vary depending on 
the wavelength since the dominant sources of the background change as a function of wavelength, on subtle 
differences in the PSF compared to a perfect Gaussian and on the clustering 
strength of the sources being stacked. For these reasons, these bias factors cannot be 
applied blindly to other far-infrared and submillimeter stacking analyses. 

We repeated the above simulations with unclustered, random distribution of the 24$\,\mu$m sources and found a bias factor of 
1.00$\pm$0.06. This shows that the dominant effect in estimating the stacked fluxes of sources below the detection limit
indeed arises from the clustering properties of the sources. If they were unclustered, there would not be a bias.

We also performed the above simulations without step 3 to test stacking biases without first subtracting the detections (as Devlin et al.~2009 and Marsden et al.~2009 have done). As expected, we find even higher bias factors than those listed in Table 4, 
with a value of 1.6 at 250\,$\mu$m instead of 1.16. Factoring in this bias factor at 250\,$\mu$m, we get IGL$_{corr}=4.6\pm0.4$ nW\,m$^{-2}$\,sr$^{-1}$ which is consistent with our best estimate of IGL$_{corr}=5.2\pm0.5$ nW\,m$^{-2}$\,sr$^{-1}$ from the 24$\,\mu$m galaxies.

\subsection{Cosmic Variance}

Although the submillimeter stacking can only be undertaken with the deep BLAST observations taken in the
ECDF-S, we can compare the stacking values at 70\,$\mu$m between the GOODS-N and ECDF-S observations to
assess if cosmic variance plays a significant role in different measures of the IGL. We choose only the
central 777\,arcmin$^{2}$ of ECDFS which has 70\,$\mu$m coverage of at least half its peak exposure time.
Using a procedure identical to that described earlier in Section 2.1.1, we find that the 179 individually
detected $>6\sigma$ sources in the 70\,$\mu$m map contribute 1.72\,\cirbu. Stacking on the remaining 6153
24\,$\mu$m $>5\sigma$ sources in this region, down to a flux limit of 25\,$\mu$Jy, yields a stacked intensity
of 1.2$\pm$0.4\,\cirbu. After the factor of 1.15 calibration correction, we find a total IGL intensity in ECDF-S of
3.1$\pm$0.4\,\cirbu. If we had done the stacking without masking, we would have obtained a value of 2.4\,\cirbu\ for
the stacked intensity from the individually undetected sources, similar 
to the factor of 2 bias that we obtained in GOODS-N.
Although there seems to be cosmic variance between the fields which results in a 26\% lower IGL in ECDF-S than in GOODS-N, 
we conclude that our method gives consistent values for the IGL in the two fields. 

\section{The Evolution of the Infrared Luminosity Function}

The evolution of the 
far-infrared luminosity function of star-forming galaxies has been measured very accurately at $z<1.3$ 
\citep{Magnelli, Caputi, LeFloch}. For each redshift, one can adopt a spectral energy distribution
for galaxies as a function of far-infrared luminosity and calculate the observed intensity at each wavelength
from galaxies in each redshift bin. The sum of these observed intensities, integrated over all redshifts and 
luminosity, is the total EBL. By integrating between $0<z<1.3$, it is
trivial to estimate what fraction of the EBL is directly attributable to galaxies at these
redshifts and if that is consistent with the fraction of EBL that is resolved through stacking. 
We neglect the contribution of active galactic nuclei (AGN) since it has been demonstrated that
their net contribution to the EBL is $<10$\% \citep{Treister}.

The local IR LF is commonly represented as a double power-law \citep{Sanders03}. This LF has a shape of:
\begin{eqnarray}
\phi(L) & = & \phi_0 \times \left(\frac{L}{L_0}\right) ^{-0.6} : { L<L_0 } \\
        & = & \phi_0 \times \left(\frac{L}{L_0}\right) ^{-2.15}  : { L>L_0 }
\end{eqnarray}

\noindent where $L_0=10^{10.5}$\,L$_{\sun}$ and $\phi_0=10^{-2.5}$\,Mpc$^{-3}$.
Our fiducial luminosity function extends down to 10$^{7}$\,L$_{\sun}$. 

\citet{Magnelli} use the deep 24 and 70\,$\mu$m imaging data in the 
GOODS/FIDEL fields, in conjunction with spectroscopic and photometric redshifts 
to estimate the redshift evolution of the 24$\mu$m/(1+z) and 70$\mu$m/(1+z) luminosity function. By using
a combination of direct detections and stacking at 70$\mu$m the luminosity function can be directly measured to fainter
luminosities than before, thereby placing the first constraints on the faint end slope of the infrared
luminosity function at $z\sim1-2$.
\citet{Magnelli} fit model spectral energy distribution to
the 24$\mu$m and 70$\mu$m luminosities of galaxies to estimate the total IR luminosity function.
They have also shown that the CE01 templates provide an excellent fit to the observed
24 and 70\,$\mu$m luminosities of galaxies at these redshifts. 
They find that the IR luminosity function shows rapid luminosity evolution as (1+z)$^{3.5\pm0.3}$ between $0<z<1$
but appears to flatten at higher redshifts \citep{Magnelli09b}. 

We attempt to constrain the evolution of the infrared luminosity function at $z>1.3$ using the measured
EBL as the primary constraint.
In order to calculate the resultant EBL arising for different evolutionary histories for the infrared luminosity function,
we need to adopt a far-infrared spectral energy distribution.
CE01 provided a set of model spectral energy distribution for infrared luminous galaxies as a function of mid-infrared
luminosity. Since the mid-infrared luminosity is correlated with the FIR luminosity, we can associate these templates
with each luminosity bin in the LF to make extrapolations to the other wavelengths.

\subsection{Constraints on Dust Enshrouded Star-Formation at $z>1$: Modified CE01 Templates}

We adopt the parametric evolution of \citet{Magnelli09b} whereby $\phi_0$ evolves with redshift
as $(1+z)^{\alpha_D}$ while $L_0$ evolves with redshift as $(1+z)^{\alpha_L}$.
$\alpha_L$=3.5 between $0<z<1$ and $\alpha_D=-0.9$
between $0<z<1$. At higher redshifts, we leave the evolution as a free parameter and assess the
range of evolutionary parameters that are consistent with the FIR EBL measured by FIRAS.

We adopt the luminosity dependent CE01 SEDs at $z<1.3$. At $z>1.3$ we apply the $(1+z)^{3.5}$
luminosity
evolution observed between $0<z<1$ to the CE01 SEDs. For example, at $z>1.3$, we assume that a L$_{IR}=10^{12}$\,L$_{\sun}$ galaxy
has the SED of a $10^{12}\times2.15^{-3.5}$\,L$_{\sun}$ in the local Universe. 
The reason we use 2.15 as the multiplicative factor is because the highest redshift bin in which the evolution
is measured is $1<z<1.3$ and the mid point of the bin corresponds to a $1+z=2.15$.
So, we scale the
lower luminosity galaxy up by the ratio of the luminosities to represent a high redshift galaxy of a higher
luminosity. We refer to these as the modified CE01 templates.

\begin{figure}
\plotone{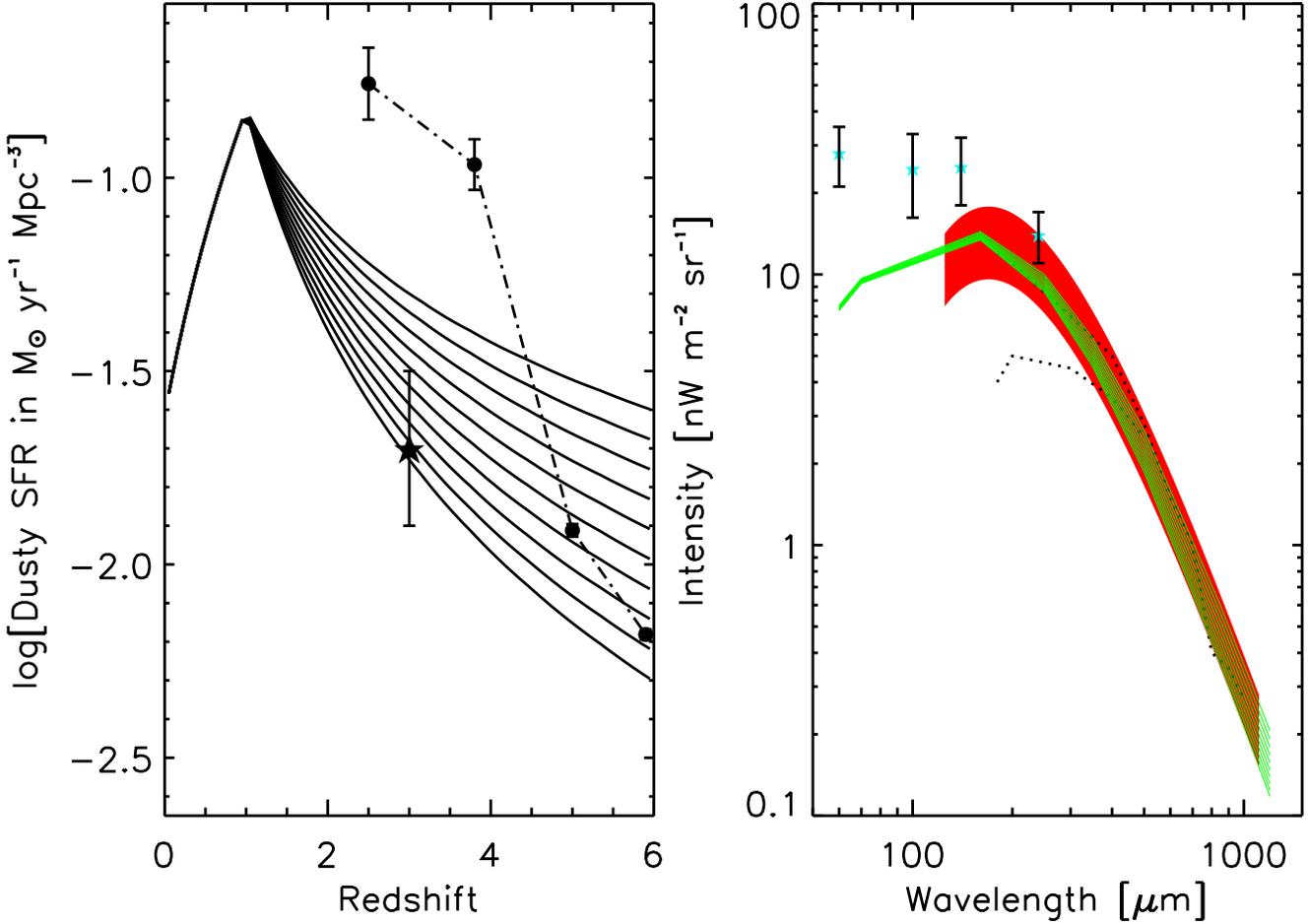}
\vspace{0.1in}
\caption{Constraints on the $z>1$ far-infrared luminosity density, and thereby dust obscured star-formation rate
from the extragalactic background light using the CE01 templates. 
The luminosity density at $z<1$ is constrained directly
by far-infrared observations presented in \citet{Magnelli}. The left panel shows different scenarios for the
evolution of the far-infrared luminosity density while the right panel shows the resultant EBL (green lines)
in comparison with observations. DIRBE estimates of the EBL are shown as blue stars while the red region
is the FIRAS estimate of the EBL. Dashed lines are the early estimates of the far-infrared EBL from
\citet{Puget}.  We allow for pure density evolution
at $z>1$, although density and luminosity evolution are degenerate as demonstrated in CE01. 
The solid black star in the left panel is the radio stacking estimate on $z\sim3$ LBGs by \citet{Carilli}.
The dot-dash 
lines and points are estimates
based on the UV-slope of galaxies from \citet{Bouwens09}.
We find that the dust obscured SFRD
cannot be as high as estimated by using the locally derived reddening prescription
on UV slope estimates \citep{Bouwens09, Reddy}.
}
\label{fig:constraint_ce01}
\end{figure}

We find that at $z>1$, if we keep $\alpha_L$=1.3, which is the tentative evolution measured at $1<z<2.3$
by \citet{Magnelli09b}, we find that $-3.2<\alpha_D<-2.3$ to avoid overproducing the FIRAS measured EBL.
This is shown in Figure \ref{fig:constraint_ce01}. We also find that the luminosity function evolution
measured at $z<1.3$ accounts for $\sim80-90$\% of the total IGL 
between $60-160$\,$\mu$m. This number drops to $\sim$67\% at 250\,$\mu$m, 55\% at 350\,$\mu$m and
44\% at 500\,$\mu$m. The statistical uncertainty on these percentages are large since the absolute level of
the background at 500\,$\mu$m is not known to better than 30\% (Table 5).

We note that there exists a discrepancy between the FIRAS EBL \citep{Fixsen} and the DIRBE measure EBL \citep{Hauser} at
$\lambda<<200$\,$\mu$m. This difference is inconsequential to the evolution at $z>1$ as can be seen by the range of
EBL values spanned by the different star-formation rate densities in Figure \ref{fig:constraint_ce01}. The evolution
at $z>1$ primarily affects the EBL values at wavelengths of $\lambda_{peak}\times(1+z)$ where $\lambda_{peak}$ is the
peak in the rest-frame far-infrared SED ($\nu$L$_{\nu}$) of galaxies. $\lambda_{peak}$ is $60-70$\,$\mu$m for 
starbursts like M82
and even for low-metallicity blue compact dwarf galaxies \citet{Engelbracht}. Thus, any effect on the EBL from
evolution between $1<z<6$ would be at $\lambda\sim 120-500$\,$\mu$m as has been discussed previously in CE01.

\begin{figure}
\plottwo{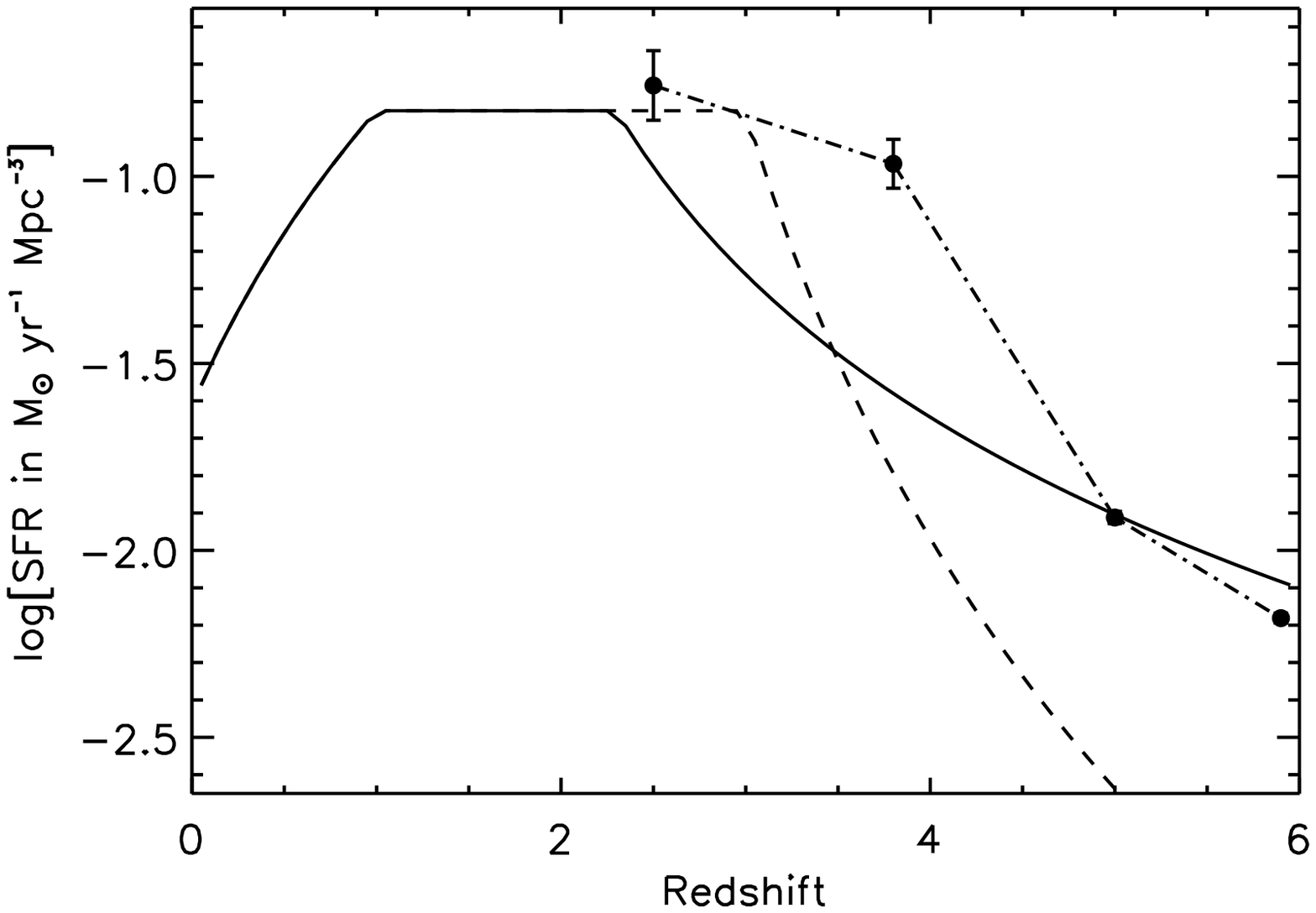}{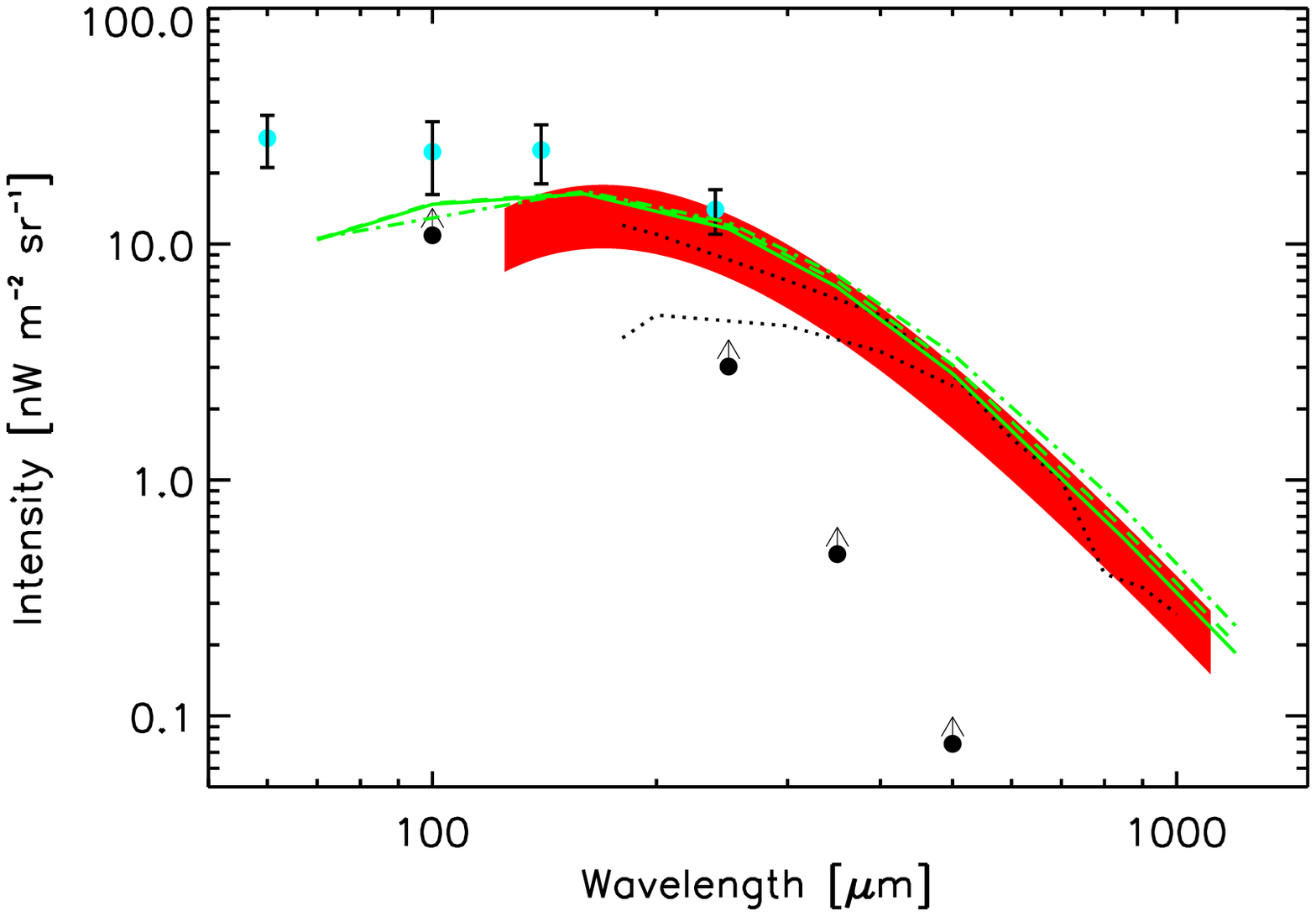}
\caption{The left panel shows three possible estimates of the 
dust obscured star-formation rate density between $0<z<6$ (See text for details). The dot-dash 
lines and points are estimates
based on the UV-slope of galaxies from \citet{Bouwens09}. The green lines in the right 
panel shows the inferred EBL if the SFR density were to evolve
as in the left panel using the CE01 templates. Linestyles are preserved in the left and right panels in that 
the solid green line in the right panels
corresponds to the solid black line in the left panel etc. The dust obscured SFRD estimated from the UV slopes
appears to over predict the EBL by $\sim$1.5$\sigma$ at $\lambda>350 \mu$m suggesting that UV slopes overestimate dust obscuration at $z>2.5$.
The solid circles show the fraction of the FIR IGL which will be directly detected with {\it Herschel}; we will need to rely on stacking to resolve the rest of the background. 
}
\label{fig:matchbouw}
\end{figure}

In an alternate scenario, we adopt the same luminosity function evolution as
\citet{Magnelli09b} at $z<1$ but we recalibrate the luminosity evolution at higher redshifts
to follow the extincted SFRD derived from the ultraviolet luminosity function of galaxies
by \citep{Bouwens09}. This is simply the difference between their extinction corrected SFRD derived using the UV slope
and the unobscured SFR from direct measurements of the UV luminosity function.
This differs from the earlier calculation at $z>1$ in that
it assumes that the IR luminosity density is flat out to $z\sim3$ and then declines monotonically between
$3<z<6$. Both these trends for the evolution of the IR LF are shown in Figure \ref{fig:matchbouw}.
We find that using the CE01 templates, we always exceed 
the FIR EBL at $\lambda>300$\,$\mu$m if we adopt a flat IR luminosity density between $1<z<2.3$ suggesting that
the UV slope extinction estimates are too high. 

%If the IR luminosity density is flat between $1<z<2.3$ in order to be consistent with the lowest redshift
%at which a dust correct SFRD is measured by \citet{Reddy} and \citet{Bouwens09}, then the decline
%of the dust obscured star-formation rate density with increasing redshift must necessarily be steeper 
%than measured from the UV to avoid violating the FIRAS measurements. The result is independent of whether
%we choose the decline to be at $z\sim2.3$ or $z\sim3$ but it cannot be at a redshift higher than 3 to avoid
%exceeding the EBL.
%This overestimate of the UV SFRD is most likely due to 
%significant contribution of nebular emission to the SED of a young stellar population, an
%effect first noticed by \citet{Reddy, Siana08}. They concluded that for galaxies with
%$<$10 Myr stellar populations, the UV slope and \citet{Meurer} relation
%overestimates the reddening partly due to the nebular emission and partly
%because the intrinsic colors of the stellar population are significantly different from the nominal
%100 Myr old stellar population template adopted as a benchmark for the intrinsic SED of a galaxy.

Another constraint at $z>1$ comes from the radio stacking analysis of \citet{Carilli}
who found that at $z\sim3$, the ratio of the radio derived star-formation rate density
to the UV-derived star-formation for the galaxies at the bright end 
of the UV luminosity function is 1.8$\pm$0.4.
This implies that the dust obscured star-formation rate density is 0.8 times the star-formation rate derived
from the observed UV values. Based on the \citet{Reddy} measure of the UV luminosity function, this translates to a
dust obscured star-formation rate of 0.02\,\sfru. Due to the relatively
shallow depth of the rest-frame UV observations in this field compared to the UDF, the fainter galaxies,
a fraction of which are likely to be the heavily dust obscured LIRGs and ULIRGs,
are missed. As a result, the derived value is most likely a lower limit. 

The range of possible constraints on the dust obscured star-formation rate density derived above
are consistent with the \citet{Carilli} measurement. We thereby adopt an $\alpha_L=1.3, \alpha_D=-3$ at $z>1$ as 
the ``best" model which is most consistent with the radio and the high redshift UV measurements of \citet{Bouwens09},
as well as the EBL constraints from FIRAS.

\subsection{Constraints on Dust Enshrouded Star-Formation at $z>1$: Cooler FIR Templates}

Although the CE01 models have been shown to fit galaxy SEDs out to $z\sim1.3$ \citep{Magnelli}, there is evidence that due to the increased
gas content of galaxies at higher redshifts, high redshift star-forming galaxy
SED are different, more typical of scaled up low-luminosity starburst
galaxies \citep{Pope08, Pope06, Rigby}. 
Another possible reason for the evolving SED might be the decreasing metallicity which might be responsible
for a change in grain size distribution.
The discrepancy between the MIR to FIR ratios
appears to be most striking at $L_{IR}>>10^{12}$\,L$_{\sun}$. Since these objects are only observed 
at $z>1.3$, it is unclear if this is a luminosity effect or a redshift effect or if there is a strong observational bias since
we can evaluate the properties of only the brightest, coolest galaxies at these redshifts \citep{Murphy09}.
We develop a revised library of SED in this paper which benefit from the following observations:

\begin{itemize}
\item An extensive set of spectroscopic observations of infrared luminous galaxies in the LIRG and ULIRG regime with IRS \citep{Pope08, Murphy09, Siana09, Siana08, Rigby, Armus07, Desai}.

\item Deep 70 and 850\,$\mu$m imaging in the GOODS/FIDEL fields which constrains the long wavelength spectral energy distribution of the most luminous sources \citep{Pope08, Murphy09}

\item Stacking results at 70\,$\mu$m as a function of 24\,$\mu$m luminosity at $z\sim1$ and $z\sim2$ \citep{Magnelli, Magnelli09b}

\end{itemize}

\begin{figure}
\plotone{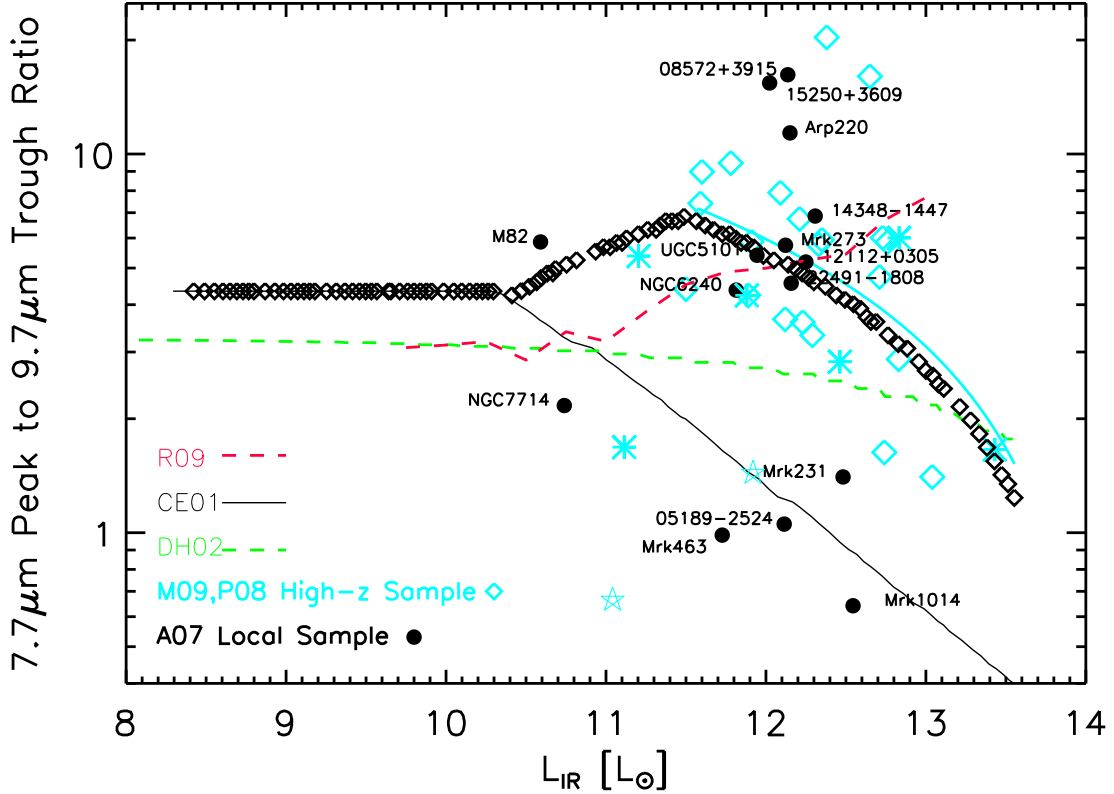}
\caption{
Ratio of the peak 7.7\,$\mu$m PAH flux to the 9$-$11\,$\mu$m flux as a function of infrared
luminosity. The ordinate is effectively a measure of the 7.7\,$\mu$m PAH equivalent width.
Due to the difficulty in measuring the absolute continuum level
in high redshift galaxies, we plot this ratio instead of an equivalent width.
The solid black circles are local ULIRGs from \citet{Armus07}, empty cyan diamonds are $0.5<z<2.5$
infrared luminous galaxies \citep{Murphy09, Pope08}, empty stars are from \citet{Siana09} and asterisks
are from \citet{Rigby}.
The solid black line shows the CE01 templates which underestimate PAH equivalent width among
the most luminous sources due to the increasing AGN contribution in local ULIRGs. The green
dashed line are the \citet{Dale} templates while the red dashed line are the \citet{Rieke09}
SED models. Neither of these models reproduce the observed trend as a result of which we
use have developed revised SEDs that follow the observational trend plotted as the black diamonds.
}
\label{mirtofir}
\end{figure}

The data seem to suggest an increase in the MIR to FIR ratio of galaxies at high luminosities such
that even the most luminous systems have large PAH equivalent width (Figure \ref{mirtofir}).
However, there are strong selection effects associated with detecting these objects at high
redshift. As a result, it is not clear if the observed galaxies span the entire range of dust temperatures
and associated far-infrared colors or are just the most extreme sources.
We note that an extensive set of spectroscopic observations of a sample of
bright 24\,$\mu$m source have been presented in \citet{Dasyra},
but these have not yet been incorporated into our database. We adopt this revised set of SED
only at $z>1.3$ since galaxies at $z<1.3$ appear to follow the CE01 templates quite well.

In \citet{Magnelli09b}, sources in the redshift bin $1.3<z<2.3$ were stacked at 70\,$\mu$m
as a function of 24\,$\mu$m/(1+z) luminosity.
It was shown that the CE01 templates overpredict the 70\,$\mu$m/(1+z) luminosity of the galaxies in
the sense that the far-infrared SEDs of these objects are too warm. 
A similar trend was found in \citet{Murphy09} and \citet{Pope08} although it was unclear if this is due to the fact that
the sample was biased by the presence of submillimeter galaxies which might have cooler dust temperatures
than average. We have modified the template
SEDs to fit the observed trend between the 24\,$\mu$m/(1+z) and 70\,$\mu$m/(1+z) luminosity of the galaxies as discussed
in Section 3.

\begin{figure}
\plotone{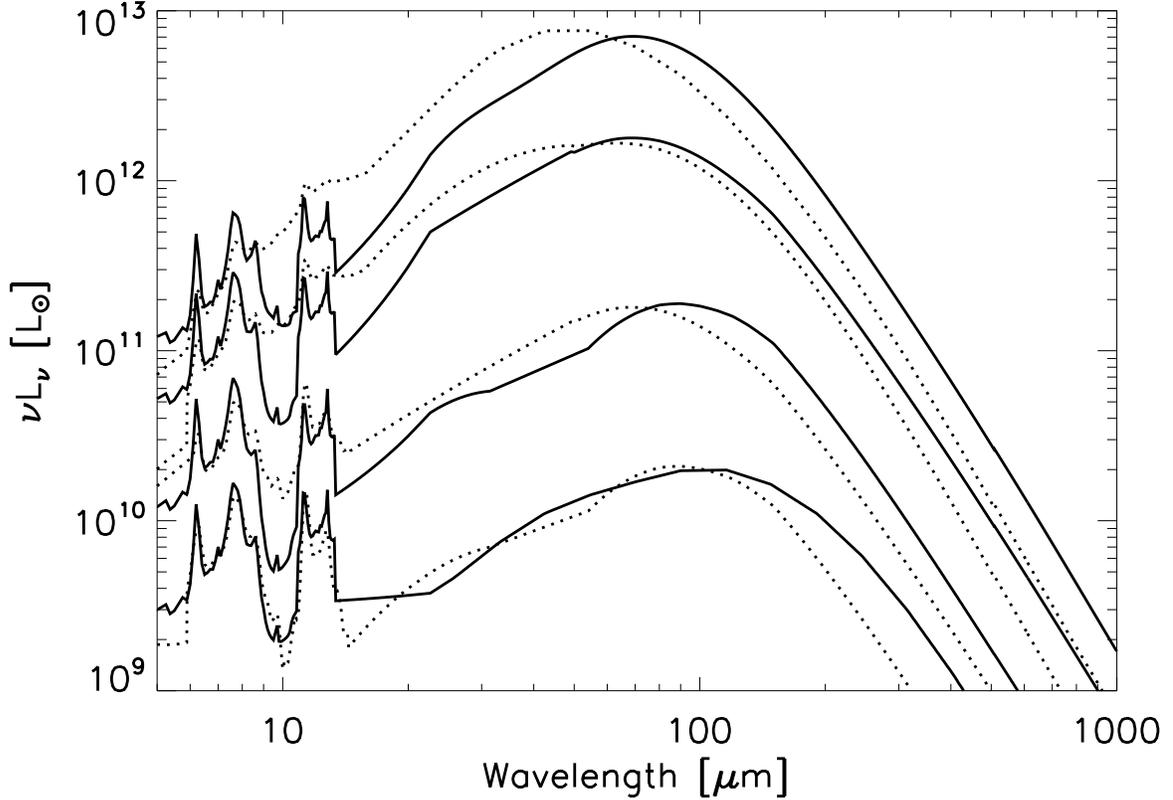}
\caption{Comparison between the SEDs developed here (solid line) to fit the mid-infrared properties
of $z>1.3$ galaxies
versus the CE01 templates (dotted line). The high redshift galaxies that have been observed appear to
show higher PAH equivalent width and cooler dust temperature which motivated the developement of the new SEDs.
However, it is unclear if this is a selection effect associated with the sensitivity of the existing far-infrared surveys.
The IR luminosities of the plotted
templates are 3.7$\times$10$^{10}$, 2.8$\times$10$^{11}$, 2.8$\times$10$^{12}$
and $10^{13}$\,L$_{\sun}$. We note that the new $10^{13}$\,L$_{\sun}$ template is consistent with the stacked
SED of submillimeter galaxies presented in \citet{Pope08}.
}
\label{fig:newsed}
\end{figure}

These new templates are shown in Figure \ref{fig:newsed} and compare favorably with the stacked
spectral energy distribution of $10^{13}$\,L$_{\sun}$ submillimeter galaxies at $z\sim2.5$ 
for which we have high quality far-infrared constraints on the SED \citep{Pope08}. We emphasize that these SED
are our first attempt to constrain the $z\sim1-2$ far-infrared SED of galaxies and further work is needed
to assess if these can be used as reliably indicators of the bolometric correction in $z>1.3$ galaxies.

We now evaluate constraints on the EBL using this set of cooler far-infrared
templates that are in Figure \ref{fig:newsed}.

\begin{figure}
\plotone{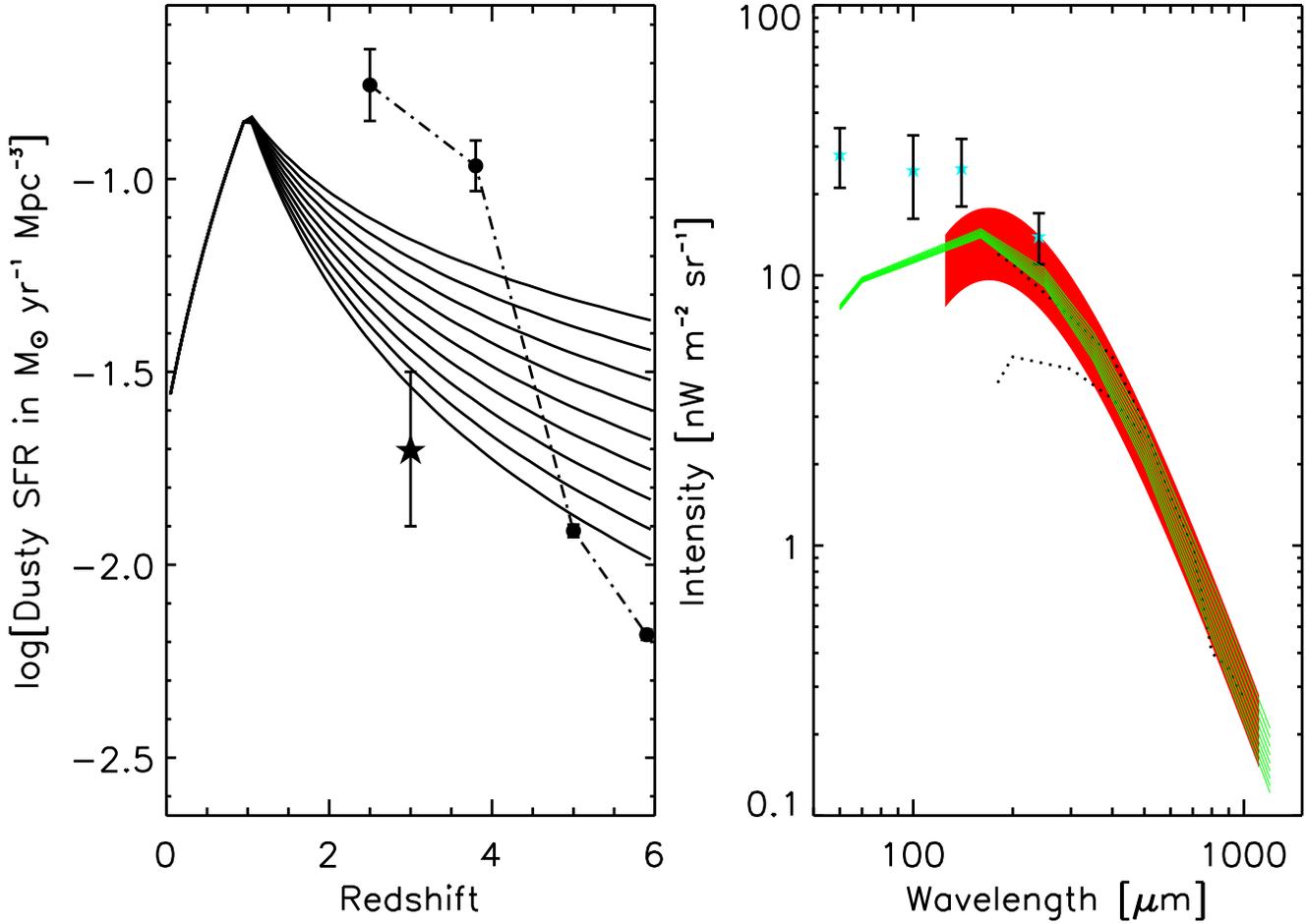}
\vspace{0.1in}
\caption{Same as Figure \ref{fig:constraint_ce01} but using the new cooler set of galaxy templates
developed in this paper for objects at $z>1.3$. The decline is marginally slower than for the modified
CE01 templates.
}
\label{fig:cp09temp}
\end{figure}

We use the CE01 templates for galaxies at $z<1.3$ but for $z>1.3$, we adopt these new cooler far-infrared templates.
For $z>1$, if we keep $\alpha_L$ fixed at 1.3, then $\alpha_D$ must
lie between $-2.8$ and $-2.0$ to be consistent with the EBL. Translating this to a dust obscured star-formation
rate density using the canonical conversion factor of 1.71$\times10^{-10}$\,L$_{\sun}$ for 1 \sfru,
we derive the upper limits on the dust obscured SFR density 
as shown in Figure \ref{fig:cp09temp}. 
A comparison with Figure \ref{fig:constraint_ce01} indicates that these values are somewhat higher than the values derived
using the CE01 templates but still overlap substantially.
The similarity between these estimates and the estimates in the previous section suggest that
the constraints on evolution at $z>1$ are not extremely sensitive to slight changes in far-infrared SED of galaxies at high
redshift.
We thereby adopt $\alpha_L=1.3, \alpha_D=-2.0$ at $z>1$ as our ``maximal"
dust obscured star-formation rate density model since it results in the highest star-formation rate density at $z>1$ without
violating the EBL values.

\begin{figure}
\plotone{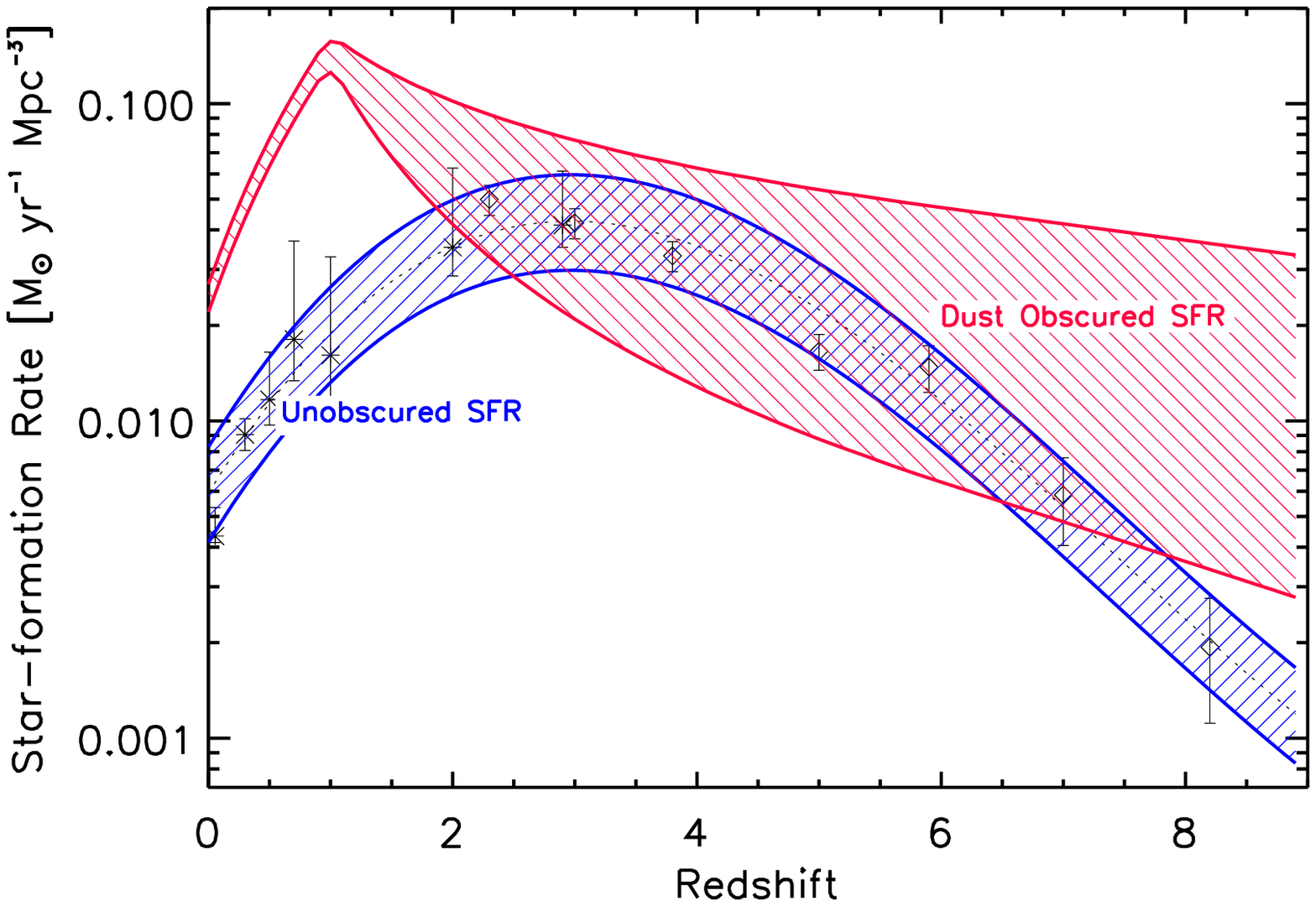}
\caption{The unobscured (blue hatched) and dust obscured (red hatched)
star-formation rate density with cosmic time. The blue dotted line is a simple polynomial fit to 
the ultraviolet measured SFR densities in \citep{Schiminovich, Reddy, Bouwens09} while the region
around it spans the range of measured uncertainties. The red region is the obscured SFR density
derived from our analysis. The constraints at $z>1$ on the obscured SFR density are ambiguous.
If the radio stacking point \citep{Carilli} is robust, then the lower limit of 
the region would be the best estimate of the obscured SFRD while the UV slope
estimates would suggest the upper line.}
\label{fig:dustyvnd}
\end{figure}

A useful consistency check on the model comes from the analysis of the 850\,$\mu$m selected galaxy population by \citet{Wall}.
They found that submillimeter galaxies at $1.8<z<4$ have a surface density of $\sim$1800$\pm$600\,deg$^{-2}$ with the surface density
at $z<1.8$ being a factor of 5 larger. Based on the derived evolution at $z>1$, we obtain a $z>1.8$ surface density of 1100\,deg$^{-2}$ for the galaxy
population with S$_{850}>2$\,mJy. Higher values for the surface density can be obtained simply by cutting off the faint end
of the infrared luminosity function at $\sim$10$^{9}$\,L$_{\sun}$.
Although the ``maximal" evolution derived here
is $\sim$2.5$\sigma$ above the radio stacking constraint of \citet{Carilli}, it indicates that the two models presented in this section and in Section 3.1
must straddle the entire range of $z>1$ evolutionary histories for the IR luminosity function.
These constraints on the dust obscured component of the star-formation rate density are compared with 
the unobscured star-formation rate density in Figure \ref{fig:dustyvnd}.

As we discussed in Section 3.1, we assess if it is possible to obtain consistency between these measurements and the UV slope
estimates using the new templates.
If the IR luminosity density is flat between $1<z<2.3$ such that it is consistent with the lowest redshift
at which a dust correct SFRD is measured by \citet{Reddy} and \citet{Bouwens09}, then the decline
of the dust obscured star-formation rate density with increasing redshift must necessarily be steeper
than measured from the UV to avoid violating the FIRAS measurements. 
We find that if the decline starts at $z=2.3$, then for $\alpha_L=0$, $\alpha_D$ must lie in the range $-6<\alpha_D<-3.4$.
This factor of $\sim$4 overestimate of the UV SFRD at $z>2.3$ is most likely due to
significant contribution of nebular emission to the SED of a young stellar population, an
effect first noticed by \citet{Reddy, Siana08}. They concluded that for galaxies with
$<$10 Myr stellar populations, the UV slope and \citet{Meurer} relation
overestimates the reddening partly due to the nebular emission and partly
because the intrinsic colors of the stellar population are significantly different from the nominal
$\sim$100 Myr old stellar population template adopted as a benchmark for the intrinsic SED of a galaxy.

We next translate our constraints on the infrared luminosity density at $z>2$ to an effective ultraviolet extinction
for comparison with the extinction derived from the UV slope.

\subsection{Evolution of Dust Extinction}

It has previously been suggested by \citet{Carilli} that the dust corrections to the UV based star-formation
rate density might be overestimates. \citet{Carilli} based their argument on the radio stacking of 
UV selected LBGs and found that at $z\sim3$, the ratio of the radio derived star-formation rate density
to the UV-derived star-formation is 1.8$\pm$0.4. If this ratio is attributed to dust extinction, then the
extinction value is a factor of $\sim2-3$ lower than previous estimates which have found extinction
corrections of $\sim$3-5.

We find that the range of values we derive for the dust obscured star-formation in the previous subsection
are in agreement with the values derived by \citet{Carilli}. Thus, we conclude that the UV slope derived
star-formation rate values are overestimates and the radio and IR are giving consistent results. Moreover
this decline in SFRD between $1<z<2.3$ is in agreement with the estimates derived from
70\,$\mu$m stacking in \citet{Magnelli09b}.

Based on these strong constraints from the EBL
on the fraction of dust obscured star-formation at $z>1$, we can
utilize the measured UV luminosity density at these redshifts to predict the average extinction.
We adopt the reddening-uncorrected UV luminosity density down to 0.04\,L$_{*,{\rm UV},z=3}$
measured by \citet{Bouwens09} between $3.5<z<6$ and those measured by \citet{Reddy} at $1.9<z<3.4$.
We adopt the ``best" constraint on the dust-obscured star-formation rate density
derived using the modified CE01 templates in section 3.1 with $\alpha_D$ at $z>1$ of $-3.0$ although
we provide uncertainties which span the range of possible FIR values derived from the analysis in Section 3.1 and 3.2.
We use the prescription of \citet{Meurer}
to convert the ratio of far-infrared to ultraviolet luminosity density to an ultraviolet extinction.
This requires converting our total infrared ($8-1000\,\mu$m) luminosity density to 
the far-infrared ($40-120$\,$\mu$m) luminosity density used
by \citet{Meurer}.
As discussed in CE01, this involves dividing the infrared luminosity by
a factor of 1.92 for the galaxies in the IRAS Bright Galaxy Sample and so we apply the same ratio here.
We have verified that this is a reasonable ratio to use even for the newer templates presented here
by integrating the far-infrared SED over the relevant wavelength ranges although the conversion
varies from $1.5-2.4$ depending on the color temperature of the far-infrared emission. 

\begin{figure}
\plotone{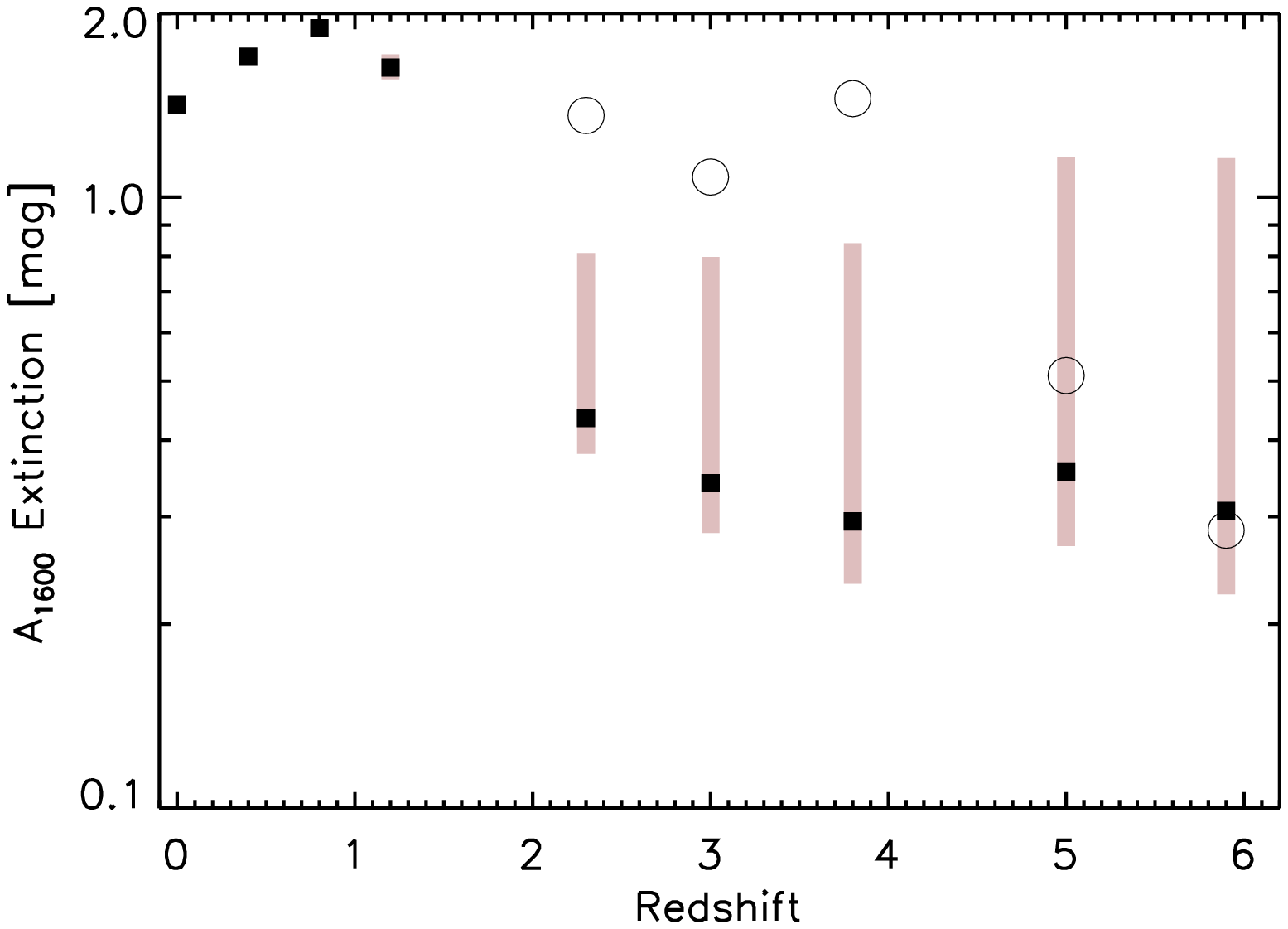}
\caption{
Our best estimate of the 1600\AA\ ultraviolet extinction (solid squares)
derived from the models presented here along with the range of values (grey regions)
that are consistent with the EBL.
Empty circles show the extinction derived from the UV slope by \citet{Reddy} and \citet{Bouwens09} which
would violate the FIRAS measured EBL estimates. The dust obscuration does increase by a factor of $\sim$2 with decreasing redshift between $2<z<6$
which is not surprising since the dust content of the Universe must be correlated with the build up of
the comoving stellar mass density which has increased by a factor of $\sim$4 between $2.3<z<5$.
}
\label{fig:a1600}
\end{figure}

Using the \citet{Meurer} conversion between the 1600\AA\ extinction and
the ratio of far-infrared luminosity to ultraviolet luminosity, we find that the upper limit to the
dust obscuration at $z>1$ imposed by the far-infrared EBL, results in extinction values that
are systematically smaller than those derived from the UV slope. The discrepancy is the largest in the $2<z<4$
range where the difference between our extinction and the UV slope extinction is a factor of $\sim$2-3 (Figure \ref{fig:a1600}).
There are two likely explanations for this. One possibility is that the increased fractional contribution due
to nebular emission, which would have a relatively red SED, is making the UV slopes steeper. The other possibility is
that the dust extinction law is changing due to the redshift evolution of dust grain sizes. For example, a Small Magellanic
Cloud extinction law would decrease the ratio of the total to unobscured star formation rate estimate for a fixed value
of the UV slope compared to the LMC or starburst extinction law which has been used to calibrate this relation (H. Shim, personal
communication).

\citet{Reddy:09} have previously argued that at $z\sim2-3$, the bulk of the star-formation occurs in faint, blue galaxies
due to the steep faint end slope of the UV luminosity function. This is certainly plausible with the constraints on
the dust obscured star-formation presented here.
However, due to the uncertainty associated with converting UV-slope
to extinction, errors are introduced in estimating dust obscured star-formation rate densities from the UV data. 
Current values derived from the UV are overestimates and should be reduced although the net effect on the comoving SFR measured
by \citet{Reddy:09} is small. 
This would be in direct contrast to previous estimates of dust obscuration at $z\sim2$, which indicate that ULIRGs dominate
the comoving star-formation rate density. 
Furthermore, this is also in sharp contrast to the estimates at $z\lesssim1.3$, where dust obscured
infrared luminous galaxies account for $\sim$65\% of the comoving SFR density \citep{Magnelli, CE01}.

We also find good evidence for there to be an increase in average extinction with decreasing redshift between $2<z<6$. 
The extinction, which is correlated with dust content increases by a factor of 2 over this redshift range. Since the
dust arises either from SNe or AGB stars \citep{Dwek:07}, the average dust content of the Universe must increase with
stellar mass density. Over this redshift range, the stellar mass density has increased by a factor of $\sim$4 suggesting
that the dust content increases as the square root of the stellar mass density.
In \citet{Chary:07}, it was shown that the average metallicity in star-forming galaxies increases
as stellar mass density to the power 0.69$\pm$0.17, over this redshift range. If metallicity were a proxy for dust content,
the observed increased in extinction that we find in this paper, is consistent with the increase in metallicity that was
found previously, indicating that the two processes are unsurprisingly co-evolving.

\subsection{Comparison between IGL Estimates from Models vs Stacking}

\begin{figure}
\epsscale{0.7}
\plotone{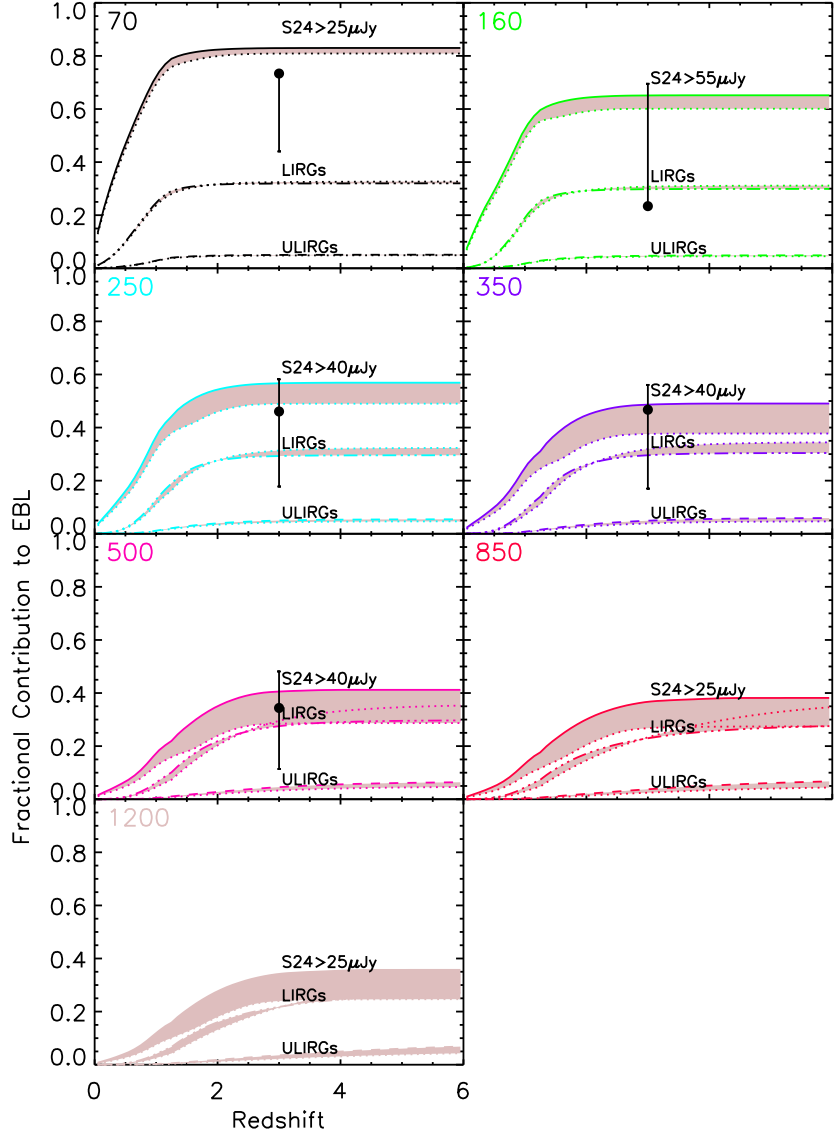}
\caption{Fractional contribution to the EBL as a function of redshift with each panel representing a different
wavelength. In each panel, we show the cumulative contribution of LIRGs, ULIRGs and galaxies above the 24\,$\mu$m
flux limit used to perform the stacking at each wavelength. For example, the GOODS-N data where the 70\,$\mu$m
stacking was done are sensitive down to $\sim$25\,$\mu$Jy while the ECDF-S data where the BLAST stacking was
done is sensitive down to $\sim$40\,$\mu$Jy.
The exact contribution of each population is dependent on the choice of FIR SED
and so we show the plausible
range for each quantity as the shaded region between the dotted and solid lines. Also plotted are our estimates
derived using stacking. The lower limit of the stacking point
is the value derived by adding the individual detections to the stacked flux from other 24\,$\mu$m sources, including
masking. The upper limit is from adding the detection to the stacked flux using no masking. These limits are not measures
of uncertainties, which are negligibly small but are provided to illustrate the difference in values if stacking is not
done carefully. The best estimate
IGL, including individual detections, stacking and with corrections for incompleteness and overcounting is plotted as 
the solid symbol. Except for 160\,$\mu$m where we cannot stack below a 24\,$\mu$m flux of 275\,$\mu$Jy due to incompleteness,
the best estimate values are in perfect agreement with the models.
}
\label{fig:lambdacontrib}
\end{figure}

We have used  deep {\it Spitzer} imaging
data at 24, 70 and 160\,$\mu$m and submillimeter imaging from 250--850$\,\mu$m to develop a reliable
stacking procedure. In particular, we have been able to correct for the biases due to the clustering around
bright detected sources in a model independent way.
We have also used semi-empirical models for the evolution of the infrared luminosity function which is constrained
by deep {\it Spitzer} imaging data at 24, 70 and 160\,$\mu$m and the COBE measured EBL. 
These two independent lines
of analysis should therefore given cross-consistent results on the estimates
derived for the IGL. Figure \ref{fig:lambdacontrib} shows the values for the IGL derived from
stacking compared with the EBL estimates derived from the model shown in Table 5.
From the models, we have also estimated the fractional contribution to the EBL from LIRGs, ULIRGs
and from bright 24\,$\mu$m sources. We have used a different 24\,$\mu$m flux limit for each panel
depending on the sensitivity of the data used for the stacking analysis.
Contributions from LIRGs and ULIRGs have been derived by integrating the luminosity function
between $10^{11-12}$\,L$_{\sun}$ and $>10^{12}$\,L$_{\sun}$ respectively. These values typically would
depend on the scatter in the far-infrared
spectral energy distribution of galaxies. For example, if a subsample of
galaxies were substantially cooler than average, they would result in a higher fraction at longer
wavelengths, but a lower fraction at shorter wavelengths.
We have therefore calculated the range
for both the modified CE01 templates described in Section 3.1 
and the newer templates presented in Section 3.2. 

We find good agreement between the stacking values and the fractional contribution inferred from the evolution of 
the infrared luminosity function which gives us confidence that our conclusions are robust. There is seemingly a
a discrepancy at 160\,$\mu$m since the stacked IGL is only $\sim$20\%, while stacking on all S$_{24}>55$\,$\mu$Jy sources
should yield $\sim$65\%. This is due to the incompleteness from masking which as shown in Section 2.1.2 implies
that we cannot stack below a 24\,$\mu$m flux density of 275\,$\mu$Jy. At a flux limit of 275\,$\mu$Jy, the model
reveals that we should obtain only about $\sim$30\% of the background, exactly what we derive from stacking.
If we instead use the IGL value of 9.8\,\cirbu\ that we get from simulating the 160\,$\mu$m map with 24\,$\mu$m sources
down to 55$\mu$Jy, we find that we would have
resolved $\sim$70\% of the EBL, in excellent agreement with the model.

For future reference, we have also plotted the fraction of the EBL that would be resolved by stacking on 850\,$\mu$m and 1.2mm maps. We note that \citet{Wang} have attempted stacking analysis
on the 850\,$\mu$m maps without subtracting the detections or masking. As discussed earlier, this results
in severe overestimates of the stacking values and thereby the derived IGL.
A stacking analysis following similar techniques presented in this paper at these longer submillimeter wavelengths will be presented in Penner et al.~(in preparation). 

\begin{figure}
\epsscale{1.0}
\plotone{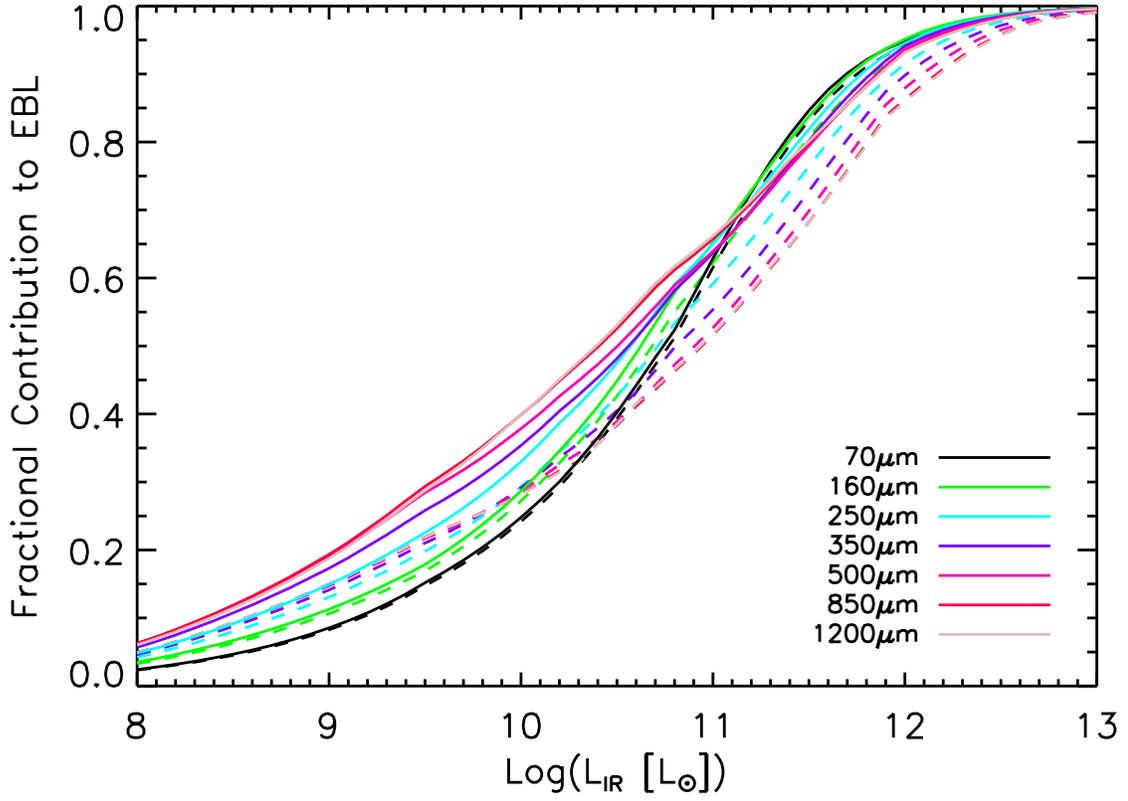}
\caption{
Fractional contribution to the EBL at different wavelengths as a function of IR luminosity, integrated over all redshifts.
The dashed lines are for the CE01 templates as discussed in Section 3.1 while the solid lines are for the new templates presented in Figure 6.
LIRGs and ULIRGs (L$_{\rm IR}>10^{11}$\,L$_{\sun}$) contribute about $\sim$35\% of the total
EBL at $\sim$70\,$\mu$m with the fraction increasing with increasing wavelength.
}
\label{fig:lumcontrib}
\end{figure}

\begin{figure}
\plotone{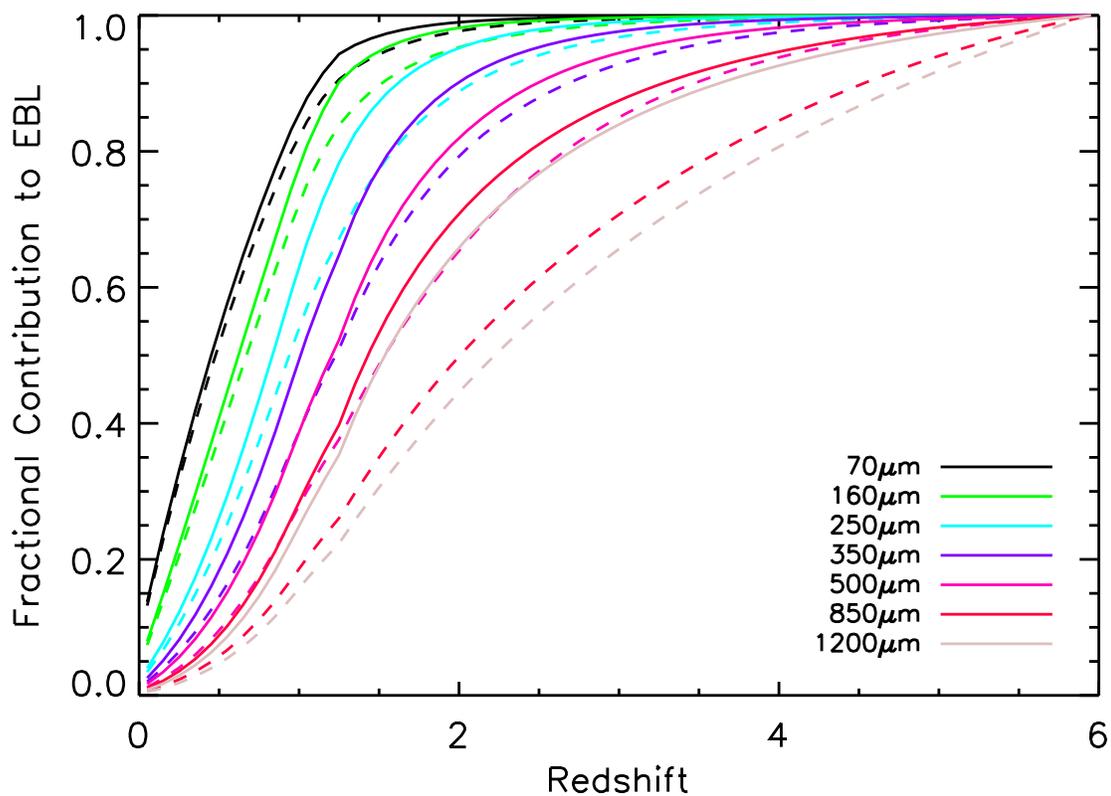}
\caption{
Fractional contribution to the EBL at different wavelengths as a function of redshift, integrated over all luminosities.
The dashed lines are for the CE01 templates while the solid lines are for the new templates presented in Figure 6.
More than $\sim$70\% of the EBL at $\lambda\lesssim500$\,$\mu$m arises from galaxies at $z<1.5$.
}
\label{fig:redshcontrib}
\end{figure}

We show in Figure \ref{fig:lumcontrib} the
fractional contribution to the EBL at different wavelengths as a function of infrared luminosity. The largest contribution
to the EBL arises from galaxies with luminosities between $10^{10-11.5}$\,L$_{\sun}$. Ultraluminous infrared galaxies
with L$_{IR}>10^{12}$\,L$_{\sun}$ contribute $\sim5-15$\% of the total EBL at any wavelength.
In Figure \ref{fig:redshcontrib}, we plot the same quantity as a function of redshift which shows that the majority of the far-IR EBL comes from galaxies at $z<1.5$.

\section{Predictions for Herschel}

Deep extragalactic surveys will be undertaken with the {\it Herschel} PACS and SPIRE instruments
between 70 and 500\,$\mu$m as part of the Guaranteed Time surveys and the Guest Observer Legacy surveys.  
Using the derived constraints on the evolution of the far-infrared luminosity function of galaxies at
$z>1$, we can make predictions for the redshift distribution of far-infrared detected galaxies
as a function of flux density and the fraction of the EBL resolved by these sources.

\subsection{Redshift Distribution}

\begin{figure}
\plotone{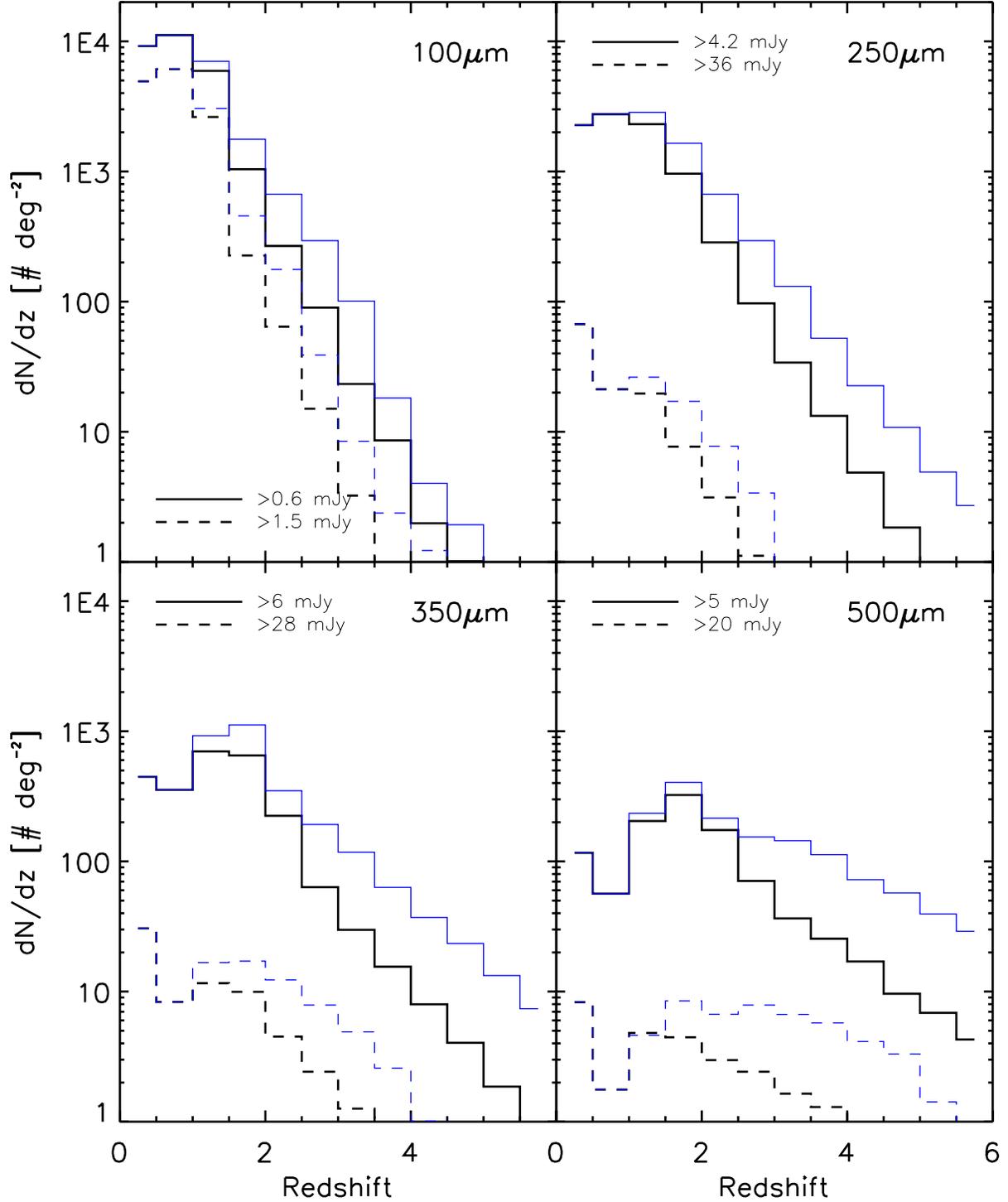}
\caption{Predicted redshift distribution of deep far-infrared surveys between 100 and 500\,$\mu$m. 
The solid lines are for a fainter flux density limit (e.g.~deep {\it Herschel} surveys such as GOODS-H) while the dashed lines are for a brighter flux density (e.g.~current BLAST limits)
at the same wavelength. The
black lines show the estimates based on the ``best" estimate of the IRLF at $z>1$ as discussed
in Section 3.1. The blue lines show the upper limits based on the ``maximal" evolution in Section 3.2
which are consistent with the FIR EBL.
}
\label{fig:dndz}
\end{figure}

We utilize both sets of templates, the modified CE01 and the new templates presented here to predict
redshift distribution of sources as a function of far-infrared flux density. As discussed in Section 3.1
and 3.2, these correspond to our ``best" model and ``maximal" model.
The redshift distributions are shown in Figure 
\ref{fig:dndz}. The deepest {\it Herschel} surveys have the potential to detect dusty ultraluminous
infrared galaxies out to $z\sim6$ (solid curves), although the deep 500\,$\mu$m survey with BLAST has potentially detected
candidate objects in the $z\sim4-5$ regime (dashed curves). Cross-identification and spectroscopic confirmation of 
these sources is likely to be difficult. Beyond $z\sim3$, the PAH features get redshifted
out of the 24\,$\mu$m bandpass as a result of which the 24\,$\mu$m is a weak tracer of dusty galaxies
at the highest redshifts. 
A deep 1.4 GHz radio survey of $\sim$15\,$\mu$Jy depth will be required to identify
the counterparts of these galaxies, assuming the local radio-infrared correlation is valid at those redshifts.
However, the surface density of $z>4$ galaxies detected at 500\,$\mu$m
is likely to be low. We anticipate that there are $\sim$2-9
such sources per GOODS field which covers about 0.046\,deg$^{2}$.

Regardless, the shape of the redshift distribution of sources shows a peak around $z\sim1$--2 and then declines
rapidly. This is broadly similar to the redshift distribution presented in \citet{Valiante} for {\it Herschel} surveys
which is unsurprising since their star-formation rate density is quite similar to that shown in Figure 7 and 8.
Our number counts appear to be quite similar to their estimates as well. At a 100$\mu$m flux density limit of 1.5 mJy,
\citet{Valiante} predict $>$10$^{4}$ sources per square degree while we predict 1.4$\times$10$^{4}$ deg$^{-2}$.
At longer wavelengths though, significant discrepancies creep in.
Above a 350$\mu$m flux density limit of 6 mJy, \citet{Valiante} predict $\sim$10$^{4}$ deg$^{-2}$ while we predict $2500-3500$ deg$^{-2}$
where the range of values arise from our ``best" model and ``maximal" model.
In comparison, Patanchon et al. (2009) derive a source density of $\sim$2900 deg$^{-2}$ above 6 mJy from a P(D)
analysis of the BLAST maps.
Similarly at a 500$\mu$m flux density limit of 5 mJy, \citet{Valiante} predict 5000 deg$^{-2}$ while we expect $1000-1500$ deg$^{-2}$.
Patanchon obtain $\sim$2000  deg$^{-2}$ at the same limit. 
The origin of this difference between \citet{Valiante} and our analysis
is likely because their fits pass through the upper range of values in the submillimeter
source counts since they adopted a flat evolution for the infrared luminosity density between $z\sim1-2$ while
our estimates favor a decline at $z>1$. Their estimate
should have resulted in an overproduction of the EBL although it is unclear from their Figure 16 whether
their measurement is in agreement with the FIRAS limits. 

We conclude that both our estimates for the redshift distribution are in agreement but that we require a slightly steeper evolution
for the $z>1$ dust obscured star-formation rate to ensure consistency with the FIR EBL which results in lower source
counts at the longest {\it Herschel} wavelengths. 

\subsection{Fraction of EBL Resolved}

\begin{figure}
\plotone{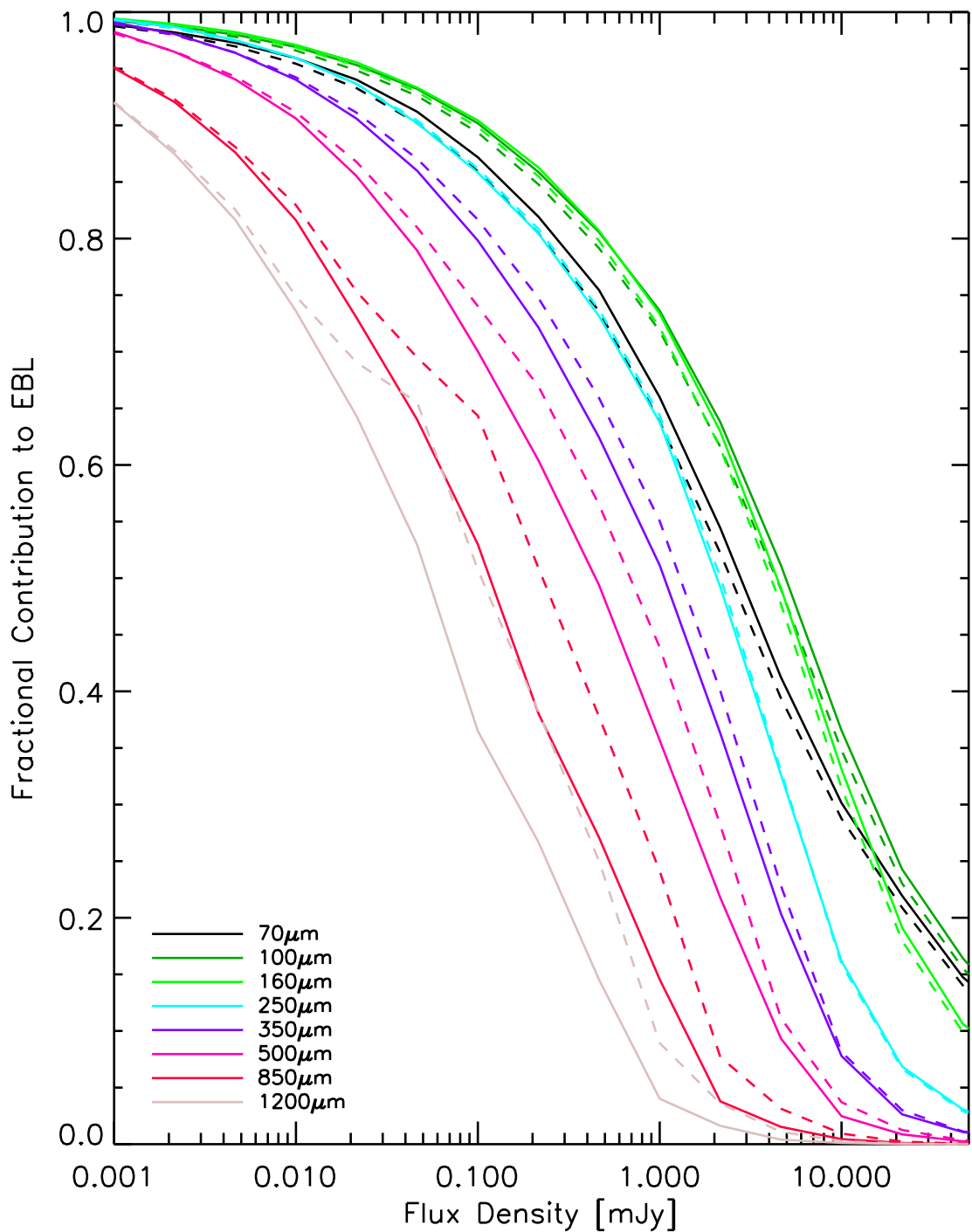}
\caption{
Fractional contribution to the EBL from sources above a specific flux density at different far-infrared
wavelengths. The solid line is for the ``best" model while the dashed line is for the ``maximal" model. The deepest surveys
with {\it Herschel} such as GOODS-H (P.I.: David Elbaz) will reach flux density limits of $\sim$1-5\,mJy.  
In order to resolve the bulk of the far-infrared/submillimeter EBL, we will have to stack hundreds of sources
to achieve stacked flux density limits more than 10 times fainter than direct detection limits.
}
\label{fig:herschelebl}
\end{figure}

Direct detections with {\it Herschel}, even with the deepest extragalactic surveys such as GOODS-H,
 will resolve only a fraction of the total FIR EBL as shown in 
Figure \ref{fig:herschelebl}. This fraction is 80\%, 48\%, 39\%, 29\% and 12\% from sources brighter
than flux density limits of 0.6, 5, 4.2, 6 and 5 mJy at the following wavelengths: 100, 170, 250, 350 and 500\,$\mu$m.
However, the superior spatial resolution of {\it Herschel} compared to either {\it Spitzer} or BLAST will alleviate
the effect of confusion noise and allow stacking to be done down to fainter limits than presented here.
As shown in Figure \ref{fig:lambdacontrib}, stacking exclusively on 24\,$\mu$m galaxy samples is unlikely
to result in significant improvement compared to the IGL estimates
presented here since the fraction of the EBL
that comes from the sample of 24\,$\mu$m selected galaxies has been estimated to be $\sim40-60$\% and
are already in agreement with our values derived from simulating the maps. Alternate possibilities such
as stacking on near-infrared selected samples of sources will have to be 
investigated while ensuring that color cuts are used to select the dusty sources from the sources that
are not actively star-forming.
We estimate that by stacking on 100s of dusty galaxies in differential flux bins identified through
their optical to near-infrared colors,
we can achieve noise values that are a factor of 10 below the direct survey limits quoted above. At these
depths, the models indicate that 95\%, 83\%, 82\%, 71\% and 68\% of the background at 100, 170, 250, 350 and 500\,$\mu$m
will be directly resolved which will be a significant improvement on the values quoted in Table \ref{tab:over}. 

\section{Implications for Zodiacal Light Models and the Near-Infrared Extragalactic Background Light}

Given the consistency between our stacking analysis and the model for the evolution of the infrared
luminosity derived based on the work by \citet{Magnelli}, we feel confident to compare the EBL
estimates at $\lambda<140$\,$\mu$m derived from DIRBE data with the model estimates. As can be seen in 
Table 5, at these wavelengths, there is a significant discrepancy between the DIRBE estimates and the models.
Constraints from the detection of TeV photons from blazars have excluded the possibility that the high
value of the EBL at 60\,$\mu$m is due to an extragalactic source population.
It is also clear that there is no unusual source population in the deep 70\,$\mu$m data that can account
for the higher value of the 60\,$\mu$m EBL. The most likely origin for the discrepancy is likely to be
the zodiacal dust cloud. By comparing our model IGL intensities with these DIRBE EBL values, we attempt to
constrain the origin of this excess emission.

The interplanetary dust cloud, or zodiacal dust, is made up of dust from asteroid and comets. The
thermal component of emission from the zodiacal cloud which dominates at $\lambda>3.5\,\mu$m and 
the sunlight scattered off the cloud at shorter wavelengths are the primary sources of foreground emission
at infrared wavelengths. Models that have been developed for the zodiacal light have utilized the
time varying component of the emission as the Earth revolves around the 
Sun \citep{Reach, Kelsall, Wright:98}. 
These three dimensional 
models broadly indicate that the dust has a temperature of $\sim$250\,K with a characteristic
distance of $\sim$1 Astronomical Unit (AU) from the Sun, although the exact values are a function of 
ecliptic latitude. Using the temporal variation of the emission however, provides poor constraints
on the isotropic component of the zodiacal emission \citep{Dwek:05}.

\citet{Wright:97} tried to constrain any additional component of zodiacal light by applying the ``very strong
no zody" condition which states that the high Galactic latitude 25\,$\mu$m intensity, where the zodiacal
emission peaks, 
should be constant and zero in one of the DIRBE observations. Although this increased the contribution
of zodiacal light emission to the diffuse intensity at both near- and far-infrared wavelengths,
the resultant extragalactic background light at near-infrared wavelengths was still 
substantially in excess of the intensity obtained by adding the light from individually detected
galaxies \citep{Levenson:08}. For example, at 3.6\,$\mu$m, corresponding to a minimum in the intensity
of the zodiacal light emission, the integrated galaxy light is thought to be $\sim5.7-9$\,\cirbu\ \citep{Fazio:04, Levenson:08}
while the EBL, using the "very strong no zody" condition is 13.3$\pm2.8$\,\cirbu \citep{Levenson:07}. The discrepancy
at shorter wavelengths i.e. 1.2 and 2.2\,$\mu$m,
is larger, which may be a result of the increasing contribution from the zodiacal
dust scattering of direct sunlight \citep{Cambresy}.

\begin{figure}
\plottwo{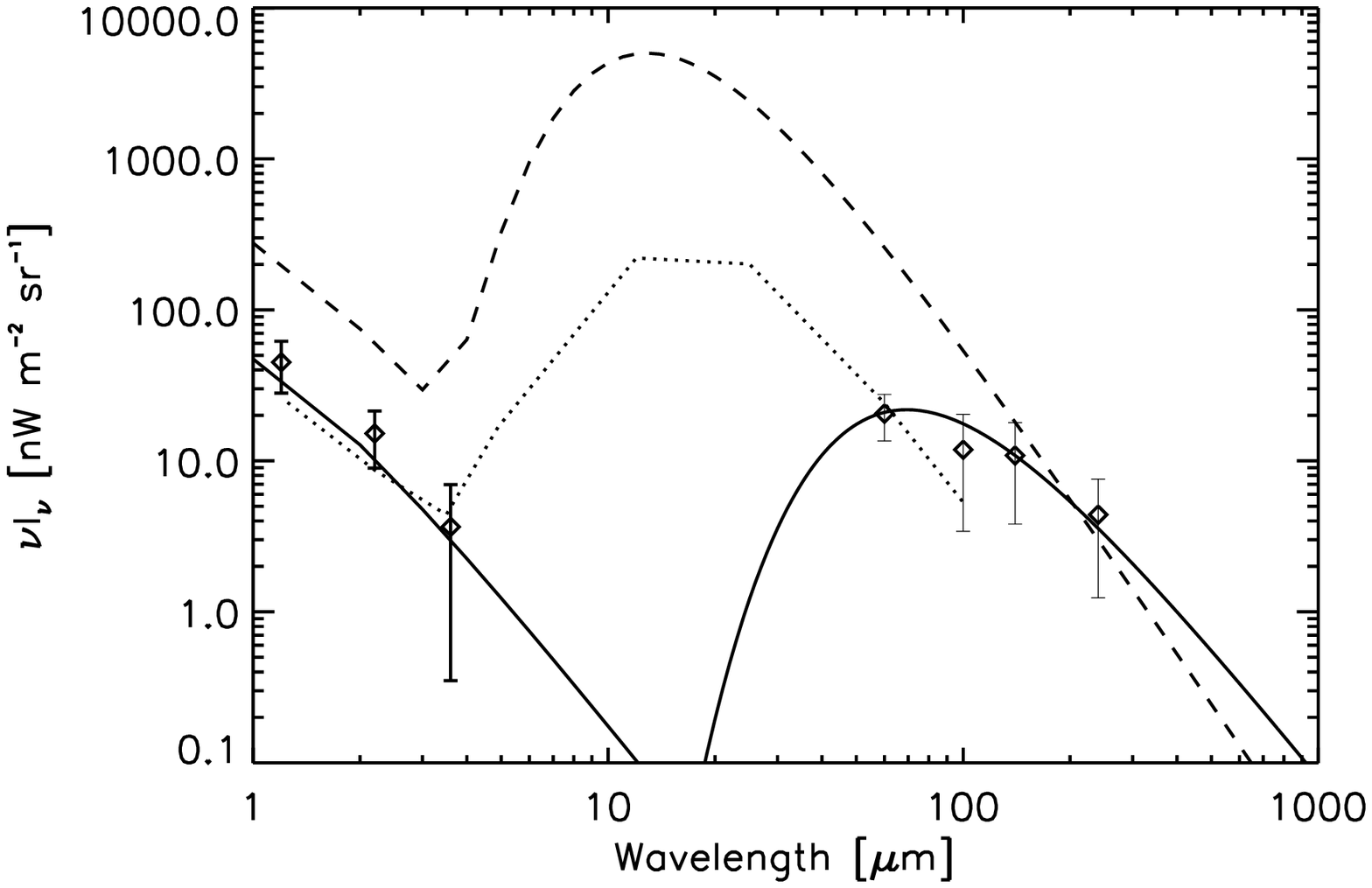}{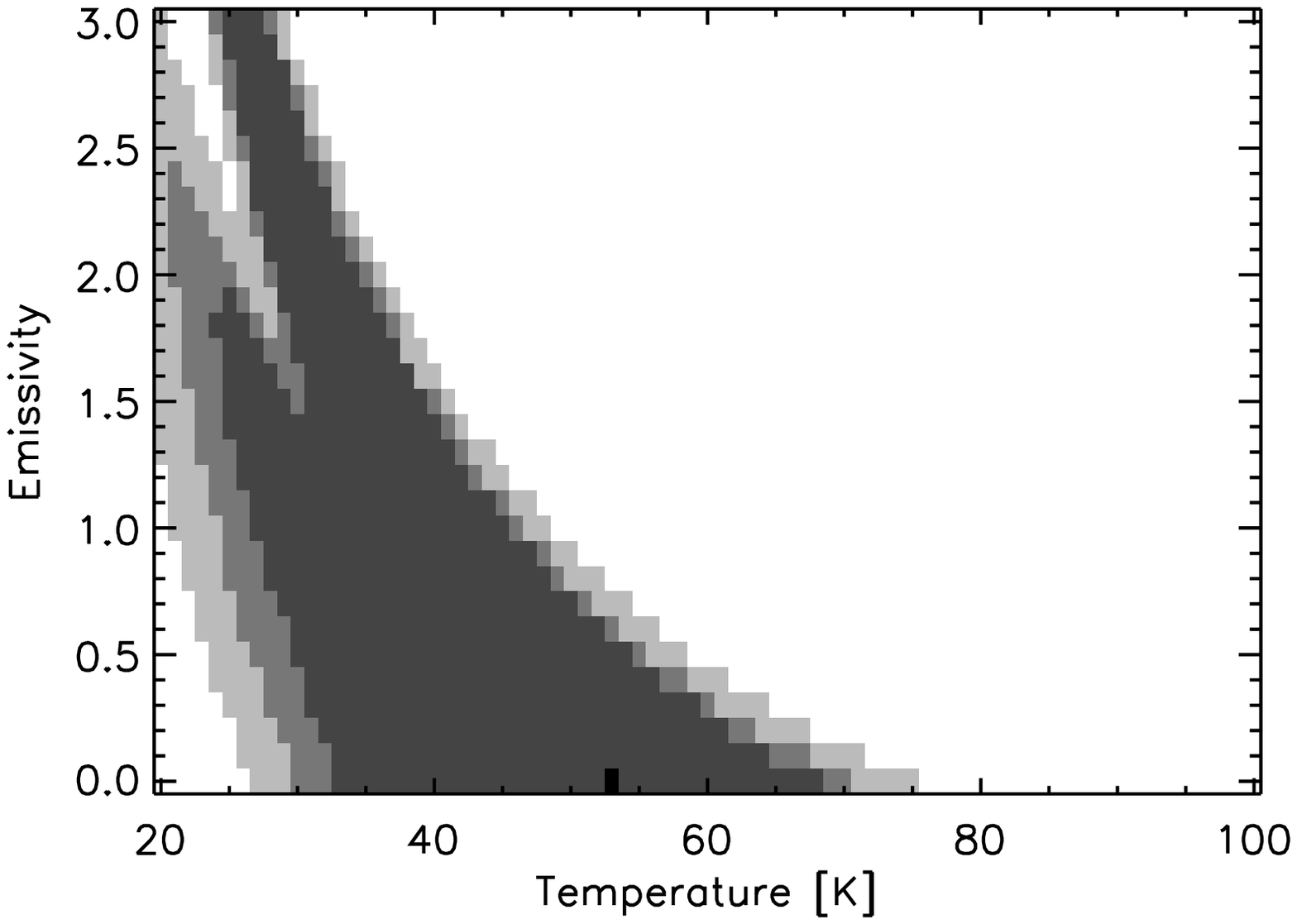}
\caption{Left panel: The symbols show the discrepancy between the integrated galaxy light and the estimated value
of the extragalactic background light at both near-infrared and far-infrared wavelengths. The solid line
shows a single component of ice mantle dust, with an albedo of $\sim$1, at a distance of $\sim$20-80 AU
that can reproduce the observed discrepancy. The dashed line shows the intensity of the zodiacal dust cloud
in GOODS-N (ecliptic latitude $\sim60\deg$) while the dotted line is the same zodiacal light spectrum scaled down by 0.1. 
The observed discrepancy between the IGL and EBL is inconsistent with
arising from the main cloud and requires a second component of emission from the outer solar system.
Right panel: Chi-square contours from fitting the far-infrared emission 
by a blackbody. The different colors show the 90, 95 and 99\% confidence levels while the best fit shown as the black square
is $\beta=0$, $T=53$\,K.
}
\label{fig:newzody}
\end{figure}

The discrepancy at near-infrared wavelengths that is discussed in the literature, combined with the difference
between the EBL and the IGL at far-infrared wavelengths suggests a common origin for the emission.
In Figure \ref{fig:newzody}, we plot the difference between the IGL and the EBL at both near- and far-
infrared wavelengths. The discrepancy is
maximum at 60\,$\mu$m, with a decreasing contribution out to 240\,$\mu$m. Although, the significance
of the discrepancy at any one wavelength varies from 2.9$\sigma$ at 60\,$\mu$m to 1.1$\sigma$ at
240\,$\mu$m, the combined intensity between 60$-$240\,$\mu$m is significant at the 3.2$\sigma$ level
after including the systematic error associated with the evolution of the far-infrared luminosity function.
As a result, this discrepancy is considered significant enough to warrant further investigation.

As shown in Figure \ref{fig:newzody},
the spectrum of this excess emission is inconsistent with the spectrum of the thermal emission in the 
zodiacal light models. The spectrum is substantially redder 
than the zodiacal emission between $60-240$\,$\mu$m. Even if we fit only the excess at 
the most significant data point, at 60\,$\mu$m,
by zodiacal emission, it falls short of explaining the difference between the IGL and EBL at both 1.2 and 2.2\,$\mu$m
although the discrepancy at each of the wavelengths is only slightly larger than 1$\sigma$.
The chi-square fit to the residual using the standard zodiacal model template
results in a $\chi^{2}\sim$300 suggesting that incorrect zodiacal light subtraction cannot be responsible
for the entirety of this residual emission.

We estimate the temperature of the blackbody corresponding to this extra component of emission.
We find that the excess emission can be well fit by a blackbody of 53\,K with a 95\% confidence interval
of 16\,K.
Due to the low significance of the discrepant emission, we do not attempt to constrain the emissivity
of the blackbody although we note that the best fit with emissivity as a free parameter is consistent
with an emissivity index of zero. The total thermal 
emission from this component is a mere 0.4\% of the main zodiacal emission in even
a dark field like the GOODS field and therefore could easily have been missed. The best fit temperature
implies that this component is most likely in the outer solar system and probably arises from
either the Kuiper Belt region or from outgassing of comets as they travel past their perihelion point
along their orbit (Chary \& Schlichting, in prep.). The existence of such a component has previously been 
suggested by \citet{Stern} although the search
for such a component in DIRBE data has previously been limited to low ecliptic latitudes
where the component is thought to be most significant \citep{Backman}.

We now assess what the albedo of this component has to be to account for the discrepancy between
the integrated galaxy light and the EBL at near-infrared wavelengths. At 1.2\,$\mu$m, we
adopt the EBL values of \citet{Cambresy}, while at 2.2 and 3.6\,$\mu$m, we adopt the EBL estimates
of \citet{Gorjian}, which are consistent with the more recent estimates of \citet{Levenson:08}.
For the integrated galaxy light, at 1.2 and 2.2\,$\mu$m, we adopt the values of \citet{Cambresy}.
At 3.6\,$\mu$m, we adopt the mean of the estimates by \citet{Fazio:04} and \citet{Levenson:08}.
Table 5 summarizes the measurements.
We find that if we assume the Kuiper Belt dust has the same grain size distribution as the zodiacal
dust cloud, the albedo of the grains has to be $\sim$4-5 times greater than the zodiacal light to reproduce the
discrepancy in the NIR background, assuming that the grains are at $\sim$ 40 AU. Since the grains
in the zodiacal dust cloud are thought to have a NIR albedo of $\sim$0.2 \citep{Kelsall, Wright:98},
the high implied albedo of $\gtrsim$0.8 for this component of emission is consistent with grains that
have an ice mantle. Remarkably, with a change in a single parameter, the albedo, we are able to simultaneously
fit the difference between the EBL and IGL at all three NIR wavelengths.

\begin{figure}
\plotone{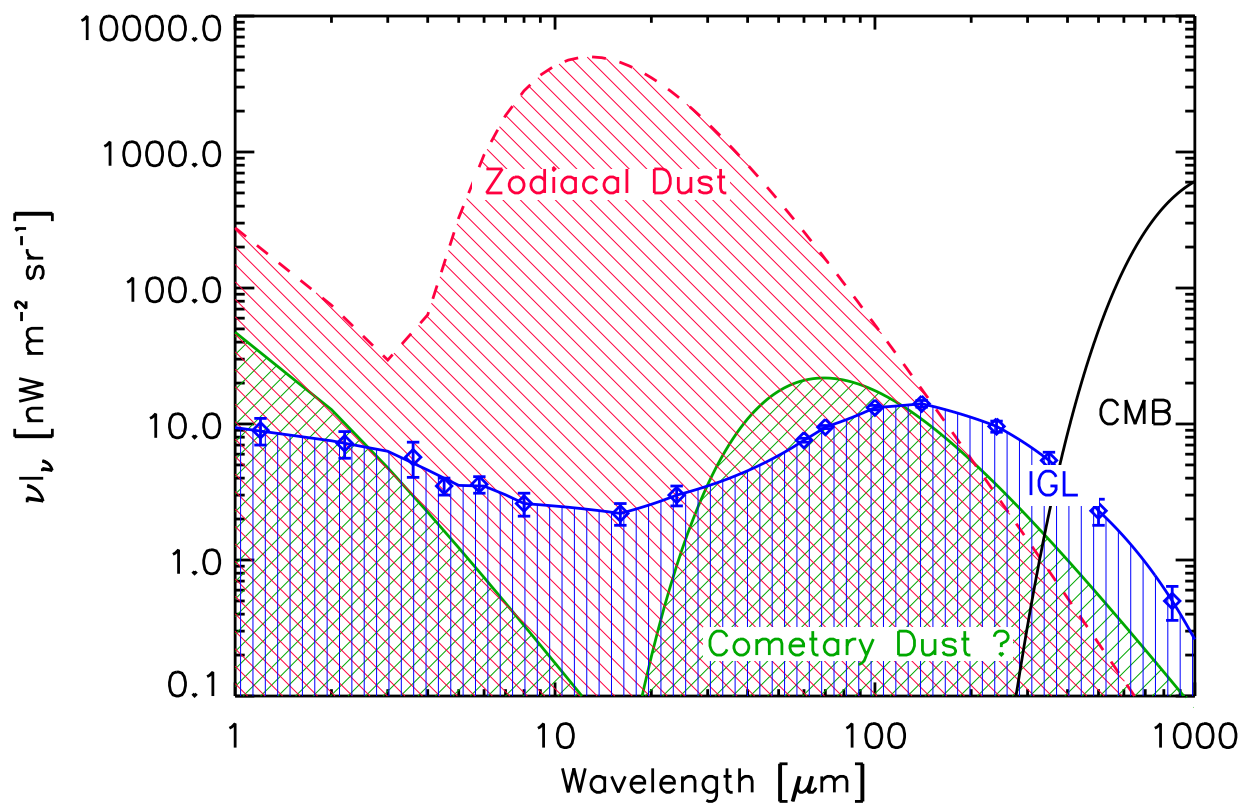}
\caption{Estimates of the relative contributions of the zodiacal dust and integrated galaxy light in
a typical extragalactic field. The latter would be isotropic and independent of choice of field.
A tentative estimate of the contribution from isotropically distributed dust in the outer solar system is also
labeled as cometary dust. Also shown is the intensity of the cosmic microwave background.
}
\label{eblcurve}
\end{figure}

We thereby conclude that there is 
evidence for the existence of a distribution of ice mantle grains at high ecliptic latitudes
at a distance of $\sim$40 AU, that is responsible for the discrepancy between the integrated light
from galaxies and the cosmic infrared background light at both near- and far-infrared wavelengths.
The total thermal emission from this component is $\gtrsim$25\,\cirbu, which is a small fraction of the 
zodiacal light emission even in high ecliptic latitude regions of sky. 
The relative contributions of the different components are shown in Figure \ref{eblcurve}.
Future work will reveal the origin and
physical properties of grains contributing to this component of emission.

\section{Conclusions}

We have utilized data from the deepest far-infrared surveys currently undertaken to evaluate the fraction
of the extragalactic background light that has been resolved into individual galaxies. 
By using a combination of direct detections
and stacking on the imaging data, 
we demonstrate that more than $\sim$50\% of the background currently remains unresolved at
all wavelengths between $100-500$\,$\mu$m. The 70\,$\mu$m IGL, after large completeness corrections
 however comprises $\sim$75\% of the extragalactic background light derived from models for the
evolution of the galaxy luminosity function.
We also demonstrate that previous estimates of the integrated galaxy light using stacking
have significantly overestimated the values by multiple counting of the flux from clustered sources and 
provide steps through which these can be corrected.

We have then, extrapolated measurements of the infrared luminosity function measured
at $z<1.2$ to higher redshifts and use
the resolved fraction of the background along with the DIRBE/FIRAS measured EBL to derive our best estimates
of the dust obscured star-formation rate at $z>1$. We find that the UV slope appears to overestimate the
dust extinction at $z>2$, most likely due to the significant contribution of nebular emission to the broadband
ultraviolet photometry. We provide average measures of the extinction at these redshifts and find that the
average extinction increases with decreasing redshift. If translated to an increase in dust content, this is consistent
with the rate of increase in metallicity with redshift, suggesting that the two processes are co-evolving.
We also quantify the
fractional contribution to the far-infrared EBL as a function of both redshift and luminosity. We find that ultraluminous
infrared galaxies contribute $\sim$5\% of the total EBL and that the bulk of the star-formation occurs in galaxies
with luminosities $\sim$10$^{10-11.5}$\,L$_{\sun}$. Finally, we use the difference between the EBL derived from our models
and the DIRBE measured values to provide tentative evidence for a contribution from ice mantle dust in the outer solar
system. This dust appears to have a total intensity of $\sim$25\,\cirbu\ at high ecliptic latitudes and can 
account for the discrepancy between the IGL and EBL at both near- and far-infrared wavelengths.

\acknowledgments
We thank Benjamin Magnelli for his work on the ECDFS 24 and 70$\mu$m catalog.
We are grateful to the BLAST team for making their data public. In particular, we thank thank Ed 
Chapin for advice on the BLAST data.
We thank Louis Levenson for helpful suggestions on the use of zodiacal light models
and William T. Reach for useful discussions regarding the use of these models.
RC is very grateful to David Elbaz and Vassilis Charmandaris for hosting his sabbatical
where the stacking analysis was first undertaken.
AP acknowledges support provided by NASA through the {\it Spitzer Space Telescope} Fellowship 
Program, through a contract issued by the Jet Propulsion Laboratory, California Institute of 
Technology under a contract with NASA.

\section{APPENDIX}

\begin{figure}
%\centering
%\includegraphics[width=3.0in]{f19.eps}
\plotone{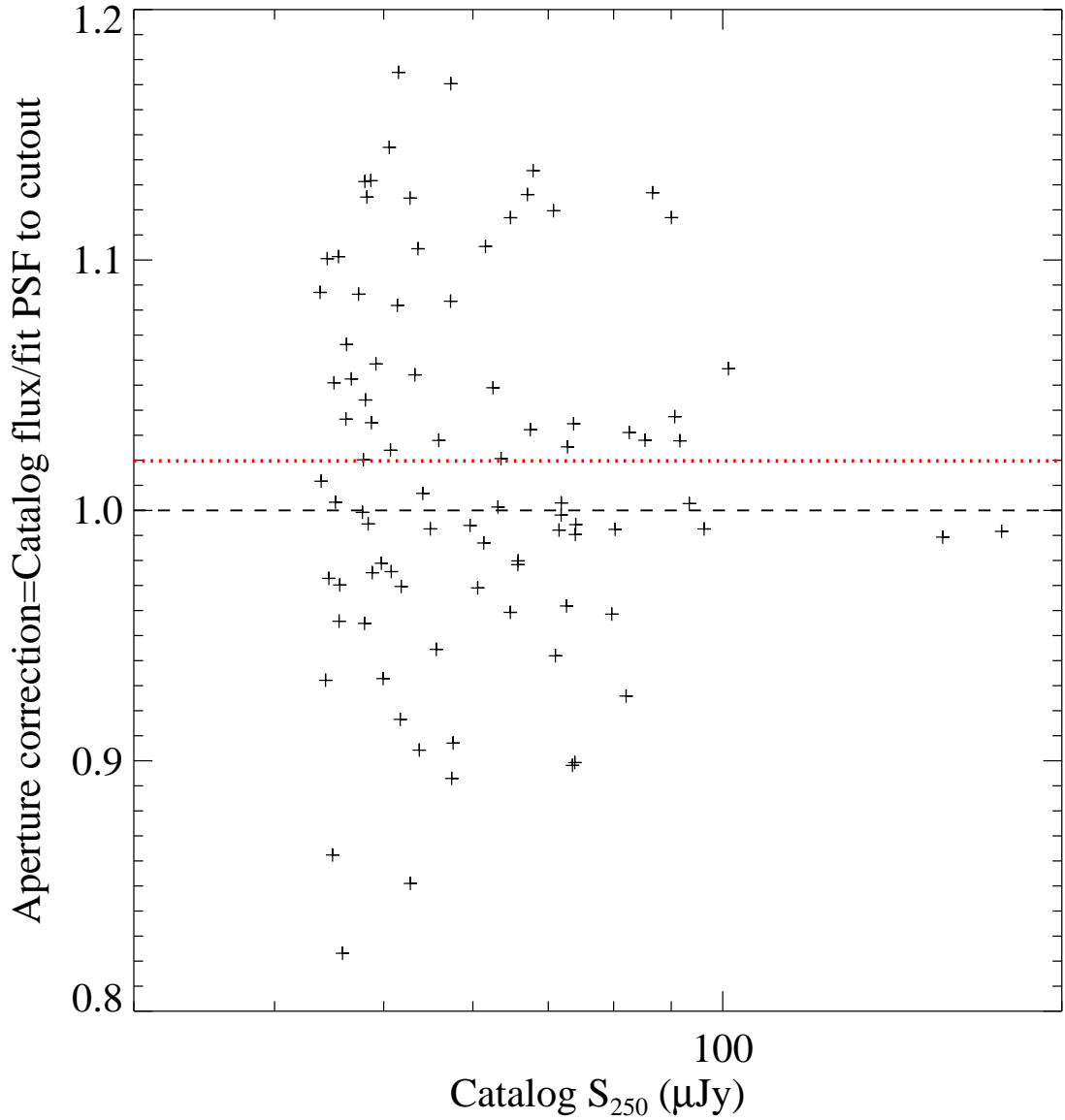}
\caption{Aperture correction for photometry on 250$\,\mu$m image cutouts. Taking the $>4\sigma$ sources in the central deep region of the BLAST ECDFS map, we compare the flux from the catalog to that measured from fitting the PSF to 9 x 9 cutouts and derive a negligible aperture correction of 1.02. At faint flux densities, the confusion from other sources increases resulting in a large scatter.}
\label{fig:apcorr}
\end{figure}

We compare the fluxes of sources measured in beam-convolved BLAST maps as measured by
\citet{Dye} with the flux measured through
aperture photometry in the unconvolved maps.
We find that the aperture flux show a larger
dispersion at fainter fluxes where the fractional contribution of other sources becomes increasingly
significant (Figure \ref{fig:apcorr}). As a result, we choose to stack the pixel flux in the beam-convolved maps 
for the BLAST data.

\begin{figure}
%\centering
%\includegraphics[width=3.0in]{f20.eps}
%\vspace{0.1in}
\plotone{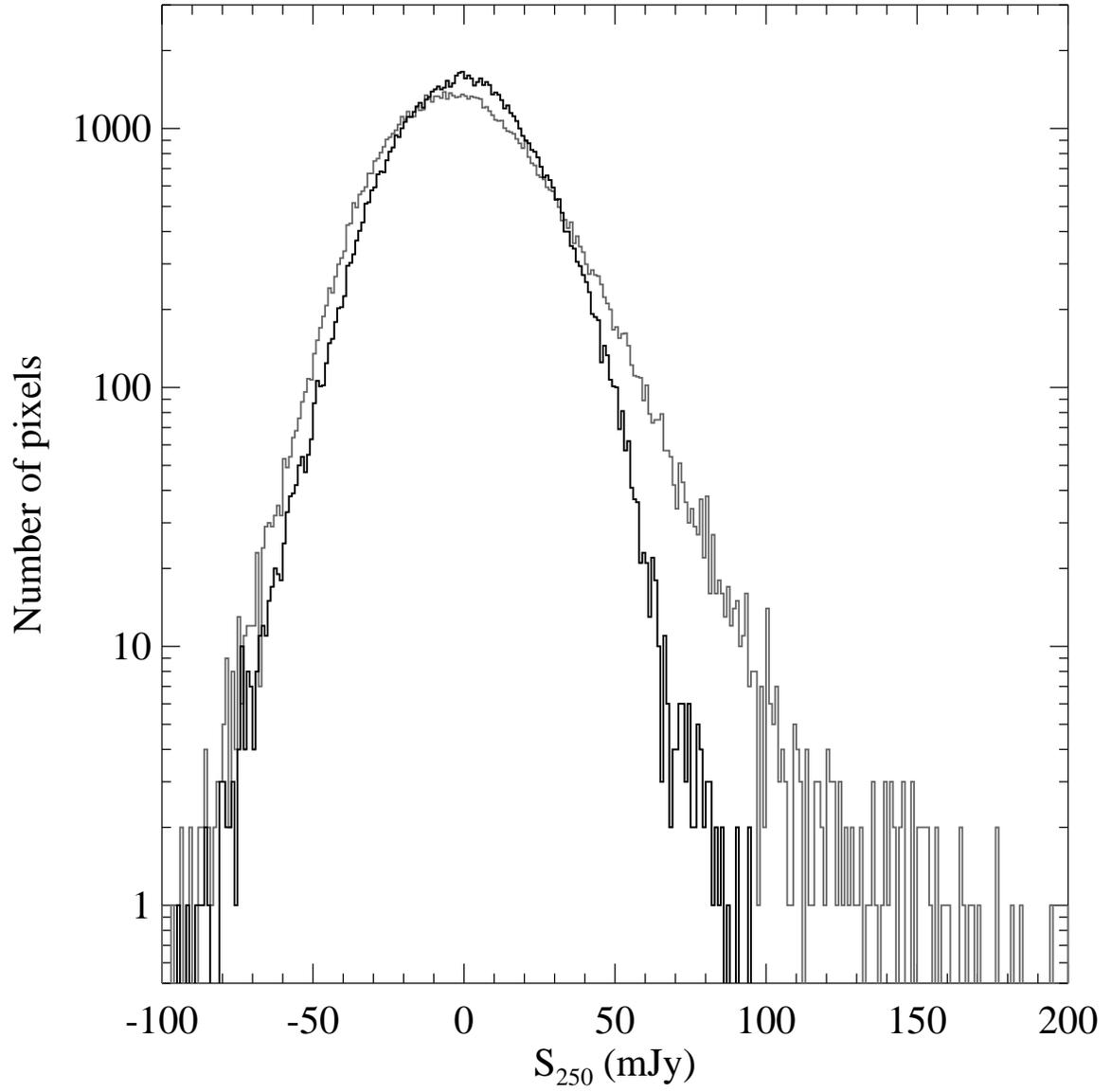}
\caption{Pixel distribution for raw (light gray) and residual (black) $250\,\mu$m signal maps. }
\label{fig:cleandist}
\end{figure}

We also compare the distribution of BLAST pixel value before and after sources are subtracted in Figure 
\ref{fig:cleandist}.
Subtracting the individually detected sources removes the positive skew in pixel values and results in a map which is
dominated by the sum of confusion and instrumental noise. We note that the confusion noise in the submillimeter
data is symmetric around zero since the beam has significant negative sidelobes while in the far-infrared data, confusion
noise will only have a positive skew.

\begin{figure}
%\centering
%\includegraphics[width=3.0in]{f21.eps}
%\vspace{0.1in}
\plotone{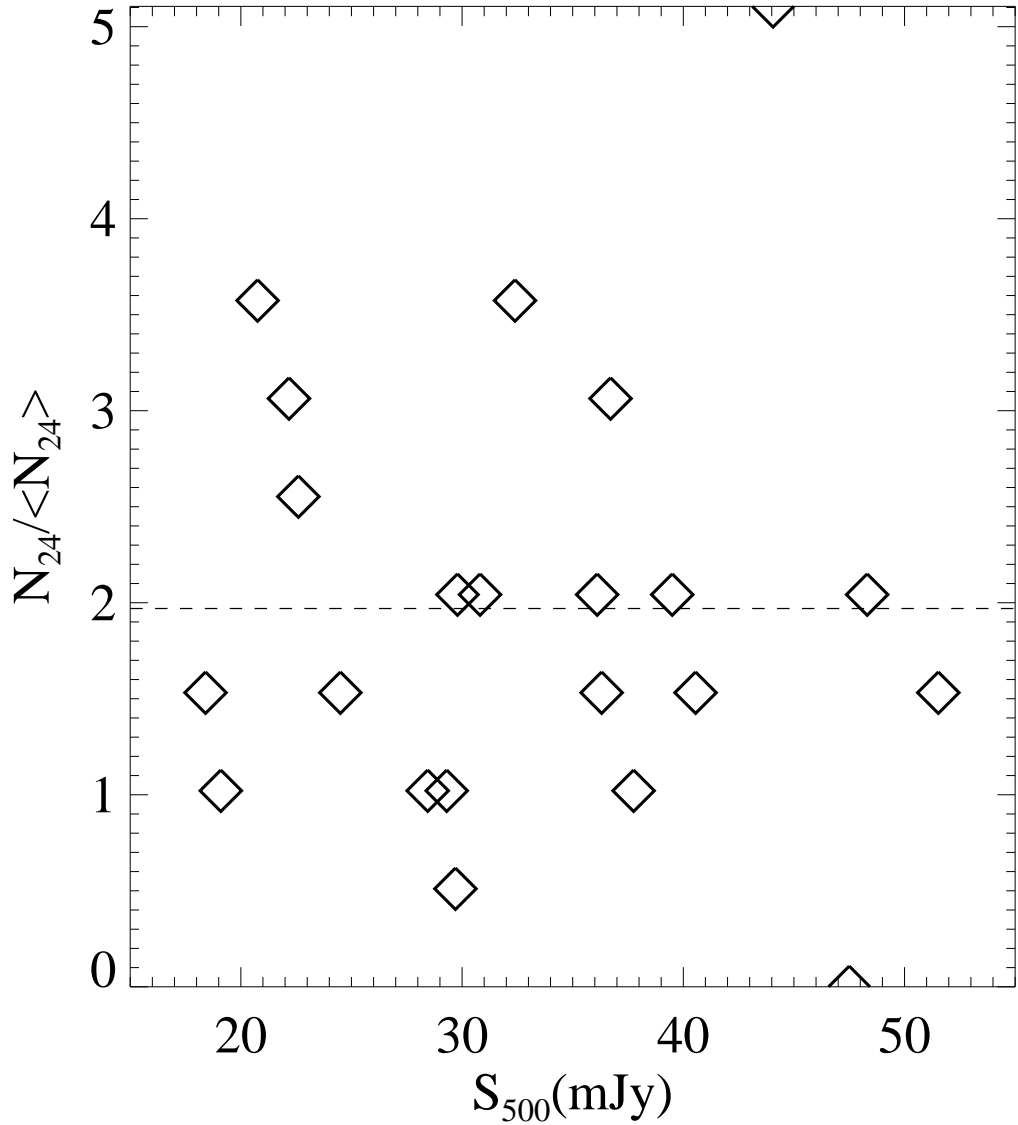}
\caption{Ratio between the number of 24\,$\mu$m sources within a radius R (see Table 1) of a BLAST 500\,$\mu$m source
and the average number of 24\,$\mu$m sources expected within the same radius assuming a completely random distribution.
The average is 2 (dashed line) and is independent of 500\,$\mu$m flux density. This implies that when stacking at
the positions of 24\,$\mu$m sources, the flux from individual 500\,$\mu$m sources will be double counted.
}
\label{fig:cluster24}
\end{figure}

We also assess how much of a role the clustering of galaxies affect the stacks by measuring the surface density of
24\,$\mu$m detected sources in the vicinity of a bright 500\,$\mu$m detected source and comparing that
to the average surface density of 24\,$\mu$m detected sources over the entire field. Figure \ref{fig:cluster24} illustrates that 
the 
bright submillimeter galaxies have a factor of two overdensity of 24\,$\mu$m detected galaxies within a radius R
compared to the 
average surface density of such sources. This suggests that the 500\,$\mu$m sources may be detected because it is the
combined flux of the individual sources. It also indicates that
when stacking at the positions of 24\,$\mu$m sources, the 500\,$\mu$m flux from the source
will thus be counted twice unless the PSF is fitted and subtracted. Stacking on maps with sources still in it
will result in a positive bias to the stacked results.

\begin{table}
\begin{center}
\caption{Description of the imaging data used for the stacking analysis.}
\vspace{0.1in}
\label{tab:blast}
\begin{tabular}{llllll}
\hline
$\lambda$ ($\mu$m) & PSF FWHM ($^{\prime \prime}$) & Pixel size ($^{\prime \prime}$) & $R$ (pixels)\tablenotemark{a} & Field & S$_{24}$/$\mu$Jy\tablenotemark{b} \\\hline
70 & 18 & 4 & 4 & GOODS-N & 25 \\
70 &    &   &     & ECDF-S & 40 \\
160 & 38 & 8 & 6.75 & EGS & 55 \\
250 & 36  & 10 & 4.5 & ECDF-S & 40 \\
350 & 42 & 10 & 5.5  & ECDF-S & 40 \\
500 & 60 & 10 & 7.5  & ECDF-S & 40 \\
\hline
\end{tabular}
\tablenotetext{a}{$R$ is the radius used for masking in the cutouts.}
\tablenotetext{b}{Flux limit of 24\,$\mu$m catalog used for stacking analysis.}
\end{center}
\end{table}

\begin{table}[h]
\begin{center}
\caption{Contribution of detected far-infrared/submillimeter sources to the IGL. }
\vspace{0.1in}
\label{tab:detected}
\begin{tabular}{llllll}
\hline
Detection level & \multicolumn{5}{c}{$\nu I_{\nu}$ [nW$\,$m$^{2}\,$sr$^{-1}$]} \\
 & $70\,\mu$m & $160\,\mu$m & $250\,\mu$m & $350\,\mu$m & $500\,\mu$m \\
\hline
%updated on May 21 by AP
$6\sigma$   & $1.9\pm0.2$ & $1.1\pm0.1$ & $0.40\pm0.01$  &  $0.232\pm0.004$   &   $0.103\pm0.002$ \\
$5\sigma$   & ... & $1.5\pm0.2$ &  $0.60\pm0.01$    &   $0.356\pm0.006$  &  $0.155\pm0.003$  \\
% $4.5\sigma$   & ... & ... & $0.76\pm0.01$  &  $0.454\pm0.007$  &    $0.180\pm0.003$ \\
$4\sigma$ & ... & $2.0\pm0.3$ & $0.97\pm0.01$   &  $0.538\pm0.008$  &    $0.199\pm0.003$ \\
% $3.5\sigma$ & ... & ... & $1.11\pm0.02$   &  $0.658\pm0.009$  & $0.223\pm0.003$  \\
% $3\sigma$   & ... & $2.6\pm0.4$ &  $1.30\pm0.02$   &  $0.757\pm0.010$  &  $0.242\pm0.003$ \\
\hline
\end{tabular}
\end{center}
\end{table}

\begin{deluxetable}{llllll}
\rotate
\tabletypesize{\scriptsize}
\tablecaption{Contribution of 24$\,\mu$m sources to the IGL as estimated from the data.  \label{tab:blast_stack}}
\tablewidth{0pt}

\tablehead{
\colhead{Method} &
\multicolumn{5}{c}{$\nu I_{\nu}$ [nW$\,$m$^{2}\,$sr$^{-1}$]\tablenotemark{b}}  \\

\colhead{} &
\colhead{$70\,\mu$m} & 
\colhead{$160\,\mu$m} & 
\colhead{$250\,\mu$m} &
\colhead{$350\,\mu$m} & 
\colhead{$500\,\mu$m}
}
\startdata
Dole et al.~2006 ($>60\mu$Jy) & 5.9$\pm$0.9 & 10.7$\pm$1.6 &   ...  & ... &  ...  \\
Devlin et al.~2009 & ... & ... &   $7.9\pm0.5$  & $4.5\pm0.3$ &  $2.0\pm0.2$  \\
\hline
No subtraction/no masking\tablenotemark{a} & 7.1$\pm$1.4 & 15.0$\pm$2.0 &  $7.30\pm0.46$ &                               $4.02\pm0.24$ &                         $1.86\pm0.14$   \\
Subtract 6$\sigma$/no masking & 5.9$\pm$1.1 (3.7) & ... & $6.70\pm0.40$ (6.30) &   $3.46\pm0.22$ (3.23) &   $1.34\pm0.11$ (1.24)  \\
Subtract 5$\sigma$/no masking & ... & 13.2$\pm$1.7 (11.8) & $6.37\pm0.36$ (5.77)  &  $3.28\pm0.20$ (2.92) &   $1.26\pm0.11$ (1.10) \\
Subtract 4$\sigma$/no masking   & ... & ... & $5.84\pm0.32$ (4.87)  &  $3.08\pm0.18$ (2.54)  &  $1.20\pm0.10$ (1.00) \\
\hline
Subtract 4$\sigma$ /masking & ... & ... &      $1.44\pm0.09$ (0.47)  &  $0.72\pm0.04$ (0.18)  &  $0.223\pm0.016$ (0.024)               \\
Subtract 5$\sigma$/masking & ... & 3.3$\pm$0.4 (1.9) &       $1.12\pm0.10$ (0.52)  &  $0.55\pm0.05$ (0.19)  &  $0.183\pm0.019$ (0.028)        \\
Subtract 6$\sigma$/masking  & 4.2$\pm$0.7 (1.96) & ... &      $0.97\pm0.11$ (0.57)  &  $0.45\pm0.05$ (0.22)  &  $0.138\pm0.021$ (0.035)        \\
Subtract 6$\sigma$/masking\tablenotemark{c}  & 7.0$\pm$1.0 (4.8) & ... &   ... &  ... &  ...       \\
No subtraction/masking & ... & ... &                              $0.71\pm0.12$ &                  $0.28\pm0.06$ &                 $0.056\pm0.029$      \\ 
\enddata
\tablenotetext{a}{These are derived by following exactly the same procedure as Dole et al. and Devlin et al. to enable a direct
comparison with their values.} 
\tablenotetext{b}{In cases where we have subtracted the detections out of the map 
we add their contribution after stacking. Values in brackets are from stacking only.}
\tablenotetext{c}{Includes a completeness correction for faint galaxies which could not be stacked on. The
completeness correction factor cannot be accurately determined for wavelengths other than 70\,$\mu$m.}
\end{deluxetable}

\begin{table}
\begin{center}
\scriptsize
\caption{Estimate of the IGL through stacking simulations.
The bias factor is the ratio of stacked flux with no masking to input flux as
determined from simulations of the residual images. Uncertainties on the bias factor reflect the range of 
templates used to convert between 24$\,\mu$m and far-infrared flux and the noise in the maps. IGL$_{\rm{corr}}$ is the
estimated contribution
 to the background from 24$\,\mu$m selected galaxies based on the simulations and
has the contribution of individually detected sources added. 
Errors on IGL$_{\rm{corr}}$ include measurement errors as well as uncertainty in the calibration. 
}
\vspace{0.1in}
\label{tab:over}
\begin{tabular}{llll}
\hline
Wavelength & Bias factor  & IGL$_{\rm{corr}}$  & fraction  \\
   &  & nW\,m$^{-2}$\,sr$^{-1}$ & \\\hline
70$\,\mu$m & $1.2\pm0.1 $  & $5.3\pm1.1$   & $\sim 56\%$  \\
160$\,\mu$m & $1.4\pm0.1 $  & $9.8\pm2.0$   & $\sim 70\%$ \\
250$\,\mu$m & $1.16\pm0.05^{\rm{a}}$  & $5.2\pm0.5$   & $\sim 54\%$ \\
350$\,\mu$m & $1.14\pm0.05$  & $2.8\pm0.3$ & $\sim 52\%$ \\
500$\,\mu$m & $1.2\pm0.06$  &  $1.0\pm0.2$  & $\sim 43\%$ \\
 \hline
\end{tabular}
\\
%\vspace{0.1in}
$^{\rm{a}}$\,The higher bias factor for 250$\,\mu$m compared to 350 microns can be explained by 
the extended wings of the PSF at 250\,$\mu$m. \\
\end{center}
\end{table}

\begin{deluxetable}{cllrr}
\rotate
\tabletypesize{\scriptsize}
\tablecaption{Estimates of the EBL and IGL}
\tablewidth{0pt}

\tablehead{
\colhead{Wavelength} &
\colhead{EBL Intensity} &
\colhead{Reference} &
\colhead{IGL Intensity} &
\colhead{Reference} \\

\colhead{$\mu$m} &
\colhead{\cirbu} &
\colhead{} &
\colhead{\cirbu} &
\colhead{} \\
}
\startdata
1.2 & 54 $\pm$ 16.8 & \citet{Cambresy} & 9 $\pm$ 2 & \citet{Cambresy} \\
2.2 & 22.4 $\pm$ 6.0 & \citet{Gorjian} & 7.2 $\pm$ 1.6 & \citet{Cambresy}\\
3.6 & 11.0 $\pm$ 3.3 & \citet{Gorjian} & 5.7$-$9.0 & \citet{Fazio:04, Levenson:08}\\
60 & 28.1 $\pm$ 7 & \citet{Finkbeiner:00} & 7.6 $\pm$ 0.25 & Model; This paper\\
70 & ... & ... & 9.5 $\pm$ 0.25 & Model; This paper\\
100 & 24.6 $\pm$ 8.4 & \citet{Finkbeiner:00} & 13.2 $\pm$ 0.5 & Model; This paper\\
140, 160 & 25 $\pm$ 7.0 & \citet{Hauser} & 14.1 $\pm$ 0.8 & Model; This paper\\
    & 12.6$\pm$4 & \citet{Fixsen} & & \\
    & 15.3$\pm$6.4 & \citet{Lagache} & & \\
240 & 14 $\pm$ 3.0 & \citet{Hauser} & 9.6 $\pm$ 0.9 & Model; This paper\\
    & 10.9$\pm$3 & \citet{Fixsen} & & \\
    & 11.4$\pm$1.9 & \citet{Lagache} & & \\
350 & 5.6 $\pm$ 1.7 & \citet{Fixsen} & 5.4 $\pm$ 0.8 & Model; This paper\\
500 & 2.4 $\pm$ 0.7 & \citet{Fixsen} & 2.3 $\pm$ 0.5 & Model; This paper\\
850 & 0.5 $\pm$ 0.15 & \citet{Fixsen} & 0.50 $\pm$ 0.14 & Model; This paper\\
1200 & 0.17 $\pm$ 0.05 & \citet{Fixsen} & 0.16 $\pm$ 0.05 & Model; This paper\\
\enddata
\end{deluxetable}

\clearpage

\end{document}